%% file: Main.tex
\documentclass[lettersize,journal]{IEEEtran}
\usepackage{amsmath,amssymb,amsfonts}
\usepackage{algorithm}
\usepackage{array}
\usepackage{algpseudocode}
\usepackage{textcomp}
\usepackage{stfloats}
\usepackage{url}
\usepackage{verbatim}
\usepackage{graphicx}
\usepackage{cite}
\usepackage{caption}
\usepackage{color,soul}
\usepackage[caption=false,font=footnotesize]{subfig}
\usepackage{float}
\usepackage{tikz}
\usepackage{pgfplots}
\usepackage{pgfplotstable}
\usepackage{pgfplots}
\pgfplotsset{compat=1.18}
\pgfplotsset{compat=1.7}
\usepackage{pgf-pie}
\usetikzlibrary{patterns}
\usepackage[dvipsnames]{xcolor}
\usetikzlibrary{plotmarks}
\usepackage{multirow}
\usepackage[caption=false,font=footnotesize]{subfig}
\usepackage{makecell} 
\usepackage{xcolor}
\usepackage{etoolbox}
\usepackage[colorlinks=true, linkcolor=blue, citecolor=blue, urlcolor=blue]{hyperref}

\makeatletter
\renewcommand{\@cite}[2]{%
  \leavevmode
  \begingroup
  \def\@temp##1{%
    \fcolorbox{green}{white}{##1}%
  }%
  [\@for\@cite@temp:=#1\do{%
    \@temp{\@cite@temp}%
  }]%
  \endgroup
}
\makeatother
\makeatletter
\let\old@ref\ref
\newcommand{\figboxref}[1]{\fcolorbox{cyan}{white}{Figure~\ref{#1}}}
\newcommand{\tabboxref}[1]{\fcolorbox{orange}{white}{Table~\ref{#1}}}
\newcommand{\algboxref}[1]{\fcolorbox{magenta}{white}{Algorithm~\ref{#1}}}

\makeatother

\begin{document}

\title{Error Understanding in Program Code: A Systematic Study of LLM-DL Combinations for Multi-label Classification}

\author{Md Faizul Ibne Amin,~\IEEEmembership{Member,~IEEE,} Yutaka Watanobe,~\IEEEmembership{Member,~IEEE,} Md. Mostafizer Rahman, Daniel M. Muepu,~\IEEEmembership{Graduate Student Member,~IEEE,} and Md. Shahajada Mia}



\maketitle

\begin{abstract}
Programming is a core skill in computer science and software engineering (SE), yet identifying and resolving code errors remains challenging for both novice and experienced developers. Errors rarely occur in isolation, which makes error understanding a multi-label problem. Large Language Models (LLMs) have shown remarkable capabilities in natural language (NL) understanding and generation tasks, but how code-specialized LLMs behave when paired with deep learning (DL) sequence decoders, and which component of such a pipeline drives performance, remains insufficiently explored. This study presents a systematic evaluation of LLM-DL combinations for multi-label error classification (MLEC) of source code. Eight fine-tuned LLMs, including CodeT5, GraphCodeBERT, CodeT5+, UniXcoder, RoBERTa, RoBERTa with a narrowed learning-rate range (RoBERTa\_LR), PLBART, and CoTexT, are integrated with GRU, LSTM, BiLSTM, and BiLSTM with an additive attention mechanism (BiLSTM-A) decoders on a real-world student-written Python code error dataset. The resulting 32 model variants, tuned with Optuna, are assessed on a comprehensive multi-label metric suite spanning average accuracy, exact match accuracy, one-error, macro and weighted precision, recall and F1-score, Hamming loss, Jaccard similarity, and ROC-AUC. In single-run evaluation, CodeT5+\_GRU performs best, with a weighted F1-score of 0.8243, average accuracy of 91.84\%, exact match accuracy of 53.78\%, Hamming loss of 0.0816, and one-error of 0.0708. To identify where this performance originates, seed-controlled baselines and component ablations are added with paired significance testing. Encoder choice has the largest effect: across four encoders sharing an identical linear classification head, the weighted F1-score spans 0.7846 to 0.8263, ordered by code specialization. On CodeT5+, the linear head exceeds the GRU hybrid under matched seeds by 0.0040 weighted F1 ($p = 0.0013$) while training about 24\% faster. Max pooling outperforms mean and attention pooling, and explicitly modeling label interactions does not improve weighted F1 despite substantial label co-occurrence. These results identify encoder quality, rather than decoder complexity, as the primary lever for MLEC and support the development of scalable automated feedback tools for programming education (PE) and SE.
\end{abstract}

\begin{IEEEkeywords}
Multi-label classification, error understanding, LLM, DL, CodeT5, GraphCodeBERT, CodeT5+, UniXcoder, RoBERTa, PLBART, CoTexT, GRU, LSTM, BiLSTM, BiLSTM-A, ablation study, programming learning, software engineering
\end{IEEEkeywords}

\section{Introduction}
\label{Introduction}
\IEEEPARstart{P}{rogramming} underpins various domains such as AI, data analysis, and modern applications, making it essential not only for Information and Communication Technology (ICT) students but also for learners across disciplines to develop computational and innovative thinking~\cite{lin2022teaching}. However, many beginners struggle with problem-solving and implementation due to the dual demands of declarative and procedural knowledge~\cite{aldalur2023improving}. Traditional programming courses often fall short because of limited duration, insufficient practice, and a lack of personalized feedback. E-learning and Online Judge (OJ) systems address this by offering extensive practice and supporting individual learning~\cite{watanobe2022online}, and many universities now employ Automated Program Assessment (APA) systems to enhance instructional effectiveness~\cite{mekterovic2020building}. Despite these advances, programmers, particularly novice students, frequently encounter errors, especially logic errors that conventional compilers cannot detect because they allow code to run but produce incorrect results. Understanding such errors remains a key challenge for both students and educators, particularly as real-world programming problems become increasingly complex~\cite{amin2024multi, shirafuji2023rule, amin2025source, wan2024automated}.

Multi-label classification (MLC) has emerged as a prominent research paradigm, in which each instance may be associated with several label categories simultaneously~\cite{bian2024takagi, liu2023robust}, and has been applied across text categorization~\cite{peng2024learning}, biomedical research~\cite{chen2022litmc}, and multimodal tasks spanning text, images, and sentiment~\cite{yu2021multi, yilmaz2021multi, zhou2023attention}. Unlike single-label settings, MLC must capture inter-label correlations and frequent co-occurrence patterns over an exponential label space~\cite{tsoumakas2008multi}. This is especially demanding in Natural Language Processing (NLP), where models encode fine-grained semantics while accounting for label dependencies~\cite{minaee2021deep}. Established approaches include problem transformation~\cite{zhang2018binary}, algorithm adaptation~\cite{hancer2025many}, neural architecture design~\cite{huang2023multi}, and ensemble strategies~\cite{moyano2018review}. Class imbalance, intricate decision surfaces, and overlapping label semantics nonetheless persist, particularly in large-scale and education-oriented applications~\cite{amin2025source, ni2025zeroed}.

Transformer-based LLMs have achieved state-of-the-art (SOTA) performance across a wide range of NLP and code-related tasks. Trained on vast corpora that include code, these models have demonstrated strong capabilities in code generation~\cite{shirafuji2023exploring}, summarization~\cite{gao2023code}, translation~\cite{10476504}, completion~\cite{izadi2024language}, program repair and classification~\cite{fatima2024flakyfix, joshi2023repair, shirafuji2023program}. Their applications now extend to educational contexts, such as aid in generating coding exercises, explaining code, and helping students understand error messages. Fine-tuning or integrating LLMs for deeper code understanding and precise error analysis nonetheless remains an active research area~\cite{ma2025large, chen2022neural, wang2025towards}. In parallel, DL models, specifically Recurrent Neural Networks (RNNs) such as GRUs, LSTMs, and BiLSTMs, model sequential patterns and token-level dependencies within source code, extracting hierarchical and temporal features relevant to program structure and semantics.

Generative LLMs (e.g., GPT) excel in reasoning-intensive tasks such as generation, translation, and repair. For classification, and particularly for MLEC, encoder-based or code-specialized LLMs are generally more suitable because they produce discriminative embeddings for downstream classification tasks. Building upon these insights, we focus on LLMs trained or adapted for programming-related and NL reasoning tasks, including CodeT5, CodeT5+, GraphCodeBERT, UniXcoder, RoBERTa, RoBERTa\_LR, PLBART, and CoTexT, which align closely with our objective. Throughout, LLM refers to these pre-trained transformer models, all of which are encoder-scale rather than generative systems. Although LLMs have proven effective in many code-related tasks, their integration with DL architectures for MLEC has not yet been systematically compared in PE and SE. This study addresses that by evaluating a unified LLM-DL architecture in which pre-trained LLM encoders are paired with recurrent decoders, namely GRU, LSTM, BiLSTM, and BiLSTM-A, across a controlled set of configurations.

In this architecture, the LLM functions as the primary encoder, transforming input code sequences into contextual representations. These representations are processed by recurrent decoders capable of modeling sequential and structural relationships in code, followed by classification layers that produce multi-label predictions. Beyond measuring the performance of these combinations, this study asks which component of the pipeline determines it, and reports controlled comparisons that isolate the contribution of the encoder, the recurrent decoder, and the pooling and label-modeling choices.
The contributions of this study are as follows.

\begin{itemize}
    \item We present a systematic study of 32 configurations pairing eight pre-trained LLM encoders with four recurrent DL decoders for MLEC of program code, tuned with Optuna and evaluated on a comprehensive multi-label metric suite over a real-world Python code error dataset.

    \item We establish seed-controlled baselines in which each encoder is paired with a simple linear classification head. Reporting mean and standard deviation across seeds, with paired significance testing, isolates the contribution of the recurrent decoder from that of the encoder.

    \item We ablate the components of the architecture under a single protocol: removing the recurrent decoder, comparing three pooling strategies, and adding a head that models interactions among labels explicitly.

    \item We characterize the dataset and the computational cost of the leading
    configurations, reporting label cardinality, density, per-label support, and
    pairwise label co-occurrence, together with training time and inference
    throughput.
\end{itemize}

Taken together, these results show that encoder choice is the primary determinant of MLEC performance. With a strong code-specialized encoder, a simple classification head matches or exceeds the recurrent hybrid at lower computational cost, while the recurrent decoder appears more relevant when encoder representations are weaker.

The rest of the article is organized as follows: Section~\ref{Related Work} reviews related studies, and Section~\ref {Motivation} presents the task description and motivation. Section~\ref{Proposed Approach} describes the integration methodology, while Sections~\ref {Experiment} and~\ref{Results} present the experimental setup and results. Section~\ref{Discussion} discusses the findings, and Section~\ref{Conclusion} concludes the study.

\section{Related Work}
\label{Related Work}
Supporting programmers in understanding, identifying, interpreting, and categorizing source code errors has become a promising area of research in both SE and PE~\cite{amin2025source}. Research~\cite{rahman2024big} analyzes large-scale coding data from an OJ system using logs, code files, and test cases to extract latent programming knowledge. Another study~\cite{watanobe2023identifying} identifies algorithms in program code using a CNN model to extract structural features. A multi-layered Bi-LSTM model with optimized hyperparameters~\cite{rahman2023multilingual} classifies multilingual source code, addressing heterogeneous codebases in PE and SE. A rule-based error classification tool~\cite{shirafuji2023rule} analyzes frequent error patterns between novice and expert programmers to categorize errors in incorrect and corrected code pairs. Additional studies address the programming learning experience more broadly~\cite{chen2022program, muepu2025comprehensive, 11151575}.

MLC has achieved strong performance across NLP domains and continues to expand into
new application areas~\cite{wu2016ml, hang2024dual}. In research~\cite{chen2022neural}, a Neural Expectation-Maximization (nEM) framework handles noisy labels in multi-label text classification by combining neural encoders with EM optimization, improving results in both single- and multi-instance settings. An Attention-Augmented Memory Network~\cite{zhou2023attention} integrates categorical memory, channel-relation, and spatial-relation modules for image MLC. Research~\cite{li2023boostxml} developed BoostXML, a DL-based extreme MLC method that improves tail-label prediction using gradient boosting, corrective alignment, and pretraining strategies. Other approaches model relationships among labels directly, exploiting high-order label correlations to improve prediction~\cite{si2023multi}. In the context of error classification, Amin et al.~\cite{amin2024multi} performed MLC by combining CodeT5 with ML-KNN, demonstrating competitive results. Further, Amin et al.~\cite{amin2025source} employed fine-tuned BERT variants (uncased and cased) for the MLEC task and compared the results with the baselines, including LLMs (CodeT5, CodeBERT) with ML classifiers (Decision Tree, Random Forest, Ensemble Learning, and ML-KNN). Several more studies in related areas can be found~\cite{wu2025graph, ye2024matchxml, mahapatra2025corrections}.

In recent years, transformer-based LLMs have become the SOTA for diverse NLP tasks~\cite{shao2024survey}. Notable models include CodeT5~\cite{wang2021codet5}, BERT~\cite{devlin2019bert}, GraphCodeBERT~\cite{guo2020graphcodebert}, CodeT5+~\cite{wang2023codet5+}, GPT~\cite{radford2018improving}, UniXcoder~\cite{guo2022unixcoder}, RoBERTa~\cite{liu2019roberta}, PLBART~\cite{ahmad2021unified}, CodeLLama~\cite{roziere2023code}, and CoTexT~\cite{phan2021cotext}, which have shown significant superiority in programming and SE-related tasks. In parallel, recurrent models such as GRU, LSTM, and BiLSTM~\cite{hochreiter1997long} and the attention mechanisms that followed them~\cite{bahdanau2014neural, vaswani2017attention} underpin much of the sequential modeling used in syntax-dependent tasks. Research~\cite{mcnichols2023algebra} proposes an LLM-based approach for algebra error classification in student responses, outperforming rule-based and syntax-tree-dependent methods. MWPES-300K~\cite{sun2025error} is a large-scale dataset of 304K math reasoning errors from 15 LLMs, paired with a dynamic error classification framework that improves performance through error-aware prompting. A further study~\cite{abbas2024novel} presents a DL approach for binary and multiclass myocardial infarction detection, showing that data balancing significantly improves performance, with a DNN reaching competitive accuracy. More recent work on LLM and DL is presented in~\cite{zhang2023survey, chen2023transrnam, rahman2025roberta}.

Although program code understanding, error classification, and multi-label learning have each advanced considerably, their intersection remains only partly mapped. Transformer-based encoders learn contextual and semantic representations of code, while recurrent DL models capture sequential dependencies among tokens, yet the two are typically studied in isolation and their combination has not been systematically compared for MLEC. Prior work in this direction has paired code embeddings with classical classifiers~\cite{amin2024multi} and fine-tuned a general-purpose encoder for the task~\cite{amin2025source}, leaving open how code-specialized encoders behave alongside recurrent decoders, and which component of such a pipeline determines performance. This study addresses that question by evaluating a unified LLM-DL architecture across multiple code-oriented encoders and recurrent decoders, together with controlled comparisons that isolate the contribution of each component.

\section{Motivation and Task Description}
\label{Motivation}
Conventional compilers and static analysis tools are effective in identifying syntactic and type-level errors, such as undefined variables or unmatched parentheses. However, they are far less effective at uncovering \textit{semantic-level logic errors}, which allow the code to compile and run but result in incorrect or unintended behavior~\cite{rahman2020source}. For instance, incorrect loop bounds, misuse of comparison operators, or misapplied functions can lead to incorrect outputs. Let a logic error be denoted as an incorrect transformation in the semantic behavior of a program $c_i$, such that $compile(c_i) = success$, ${run}(c_i) = success$, and $output(c_i) \neq expected(c_i)$. Identifying such anomalies requires interpreting program intent, control flow, and contextual semantics, which lie beyond the reach of static, rule-based checking.

While automated assessment platforms, advanced debugging tools, and modern LLMs have expanded the scope of feedback available to programmers, a single erroneous
submission often carries several interacting error types at once, and assigning all of them demands multi-label reasoning rather than a single verdict. Earlier work approached this requirement from two directions. In~\cite{amin2024multi}, code embeddings were classified by a classical multi-label learner. In~\cite{amin2025source}, a general-purpose encoder was fine-tuned end-to-end for the task. Both established that pre-trained representations carry usable error signal. Neither examined what happens when a code-specialized encoder is followed by a decoder that models token order and structure explicitly, which is the arrangement studied here. Building such a pipeline raises a further question that the design alone cannot answer: how much of the resulting accuracy comes from the encoder, and how much from the decoder placed on top of it. The answer matters in practice as well as in principle, since deploying feedback at scale depends on knowing which part of the model the accuracy is paying for.

This study evaluates a deep neural architecture that pairs pre-trained LLMs, namely CodeT5~\cite{wang2021codet5}, GraphCodeBERT~\cite{guo2020graphcodebert},
CodeT5+~\cite{wang2023codet5+}, UniXcoder~\cite{guo2022unixcoder}, RoBERTa and RoBERTa\_LR~\cite{liu2019roberta}, PLBART~\cite{ahmad2021unified}, and
CoTexT~\cite{phan2021cotext}, with sequence learners GRU, LSTM, BiLSTM, and BiLSTM-A. The LLM encoders produce contextual representations of source code, and the recurrent layers model token-level and structural dependencies across the sequence. Section~\ref{Results} reports what each part of this arrangement contributes.

\begin{figure}[t]
  \centering
  \includegraphics[width=1\linewidth]{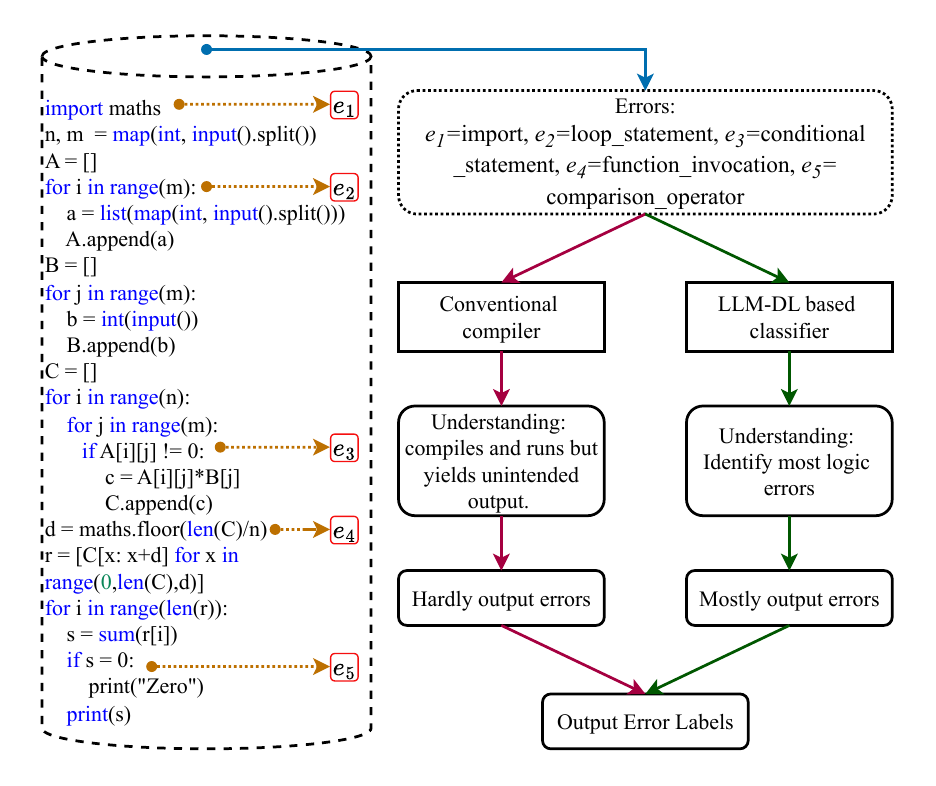}
  \caption{Source code error understanding and motivational example for the MLEC task}
  \label{ME}
\end{figure}

Let $\mathcal{D} = \{(c_i, \mathbf{y}_i)\}_{i=1}^{N}$ be a dataset consisting of $N$ code samples, where $c_i$ is a source code snippet and $\mathbf{y}_i \in \{0,1\}^L$ is a binary indicator vector over $L$ possible error types. The goal of this study is to learn a function $f: \mathcal{C} \to \mathbb{R}^L$ that maps a code snippet $c_i$ to a logit vector $\mathbf{z}_i = f(c_i)$. The predicted probability vector is then obtained as $\hat{\mathbf{y}}_i = \sigma(\mathbf{z}_i) \in [0,1]^L$, where each element $\hat{y}_{ij}$ denotes the probability of the $j$-th error type occurring in $c_i$. Formally, this can be framed as minimizing the following loss over the training set:
\vspace{-0.36em}
\begin{equation}
    \mathcal{L} = \frac{1}{N \cdot L} \sum_{i=1}^{N} \sum_{j=1}^{L} 
    \text{BCEWithLogits}(z_{ij}, y_{ij})
\end{equation}
where $\text{BCEWithLogits}$ denotes the binary cross-entropy loss applied to the raw logit $z_{ij}$ and the ground-truth label $y_{ij}$ for label $j$ of sample $i$. The example in~\figboxref{ME} illustrates the difficulty. The submission compiles and runs, yet carries five distinct error types that compiler diagnostics do not surface. Recovering all of them requires a model that reads semantic and structural cues rather than syntax alone.

\section{Combining LLMs with DL Models}
\label{Proposed Approach}
\begin{figure*}[t!]
  \centering
  \includegraphics[width=1\linewidth]{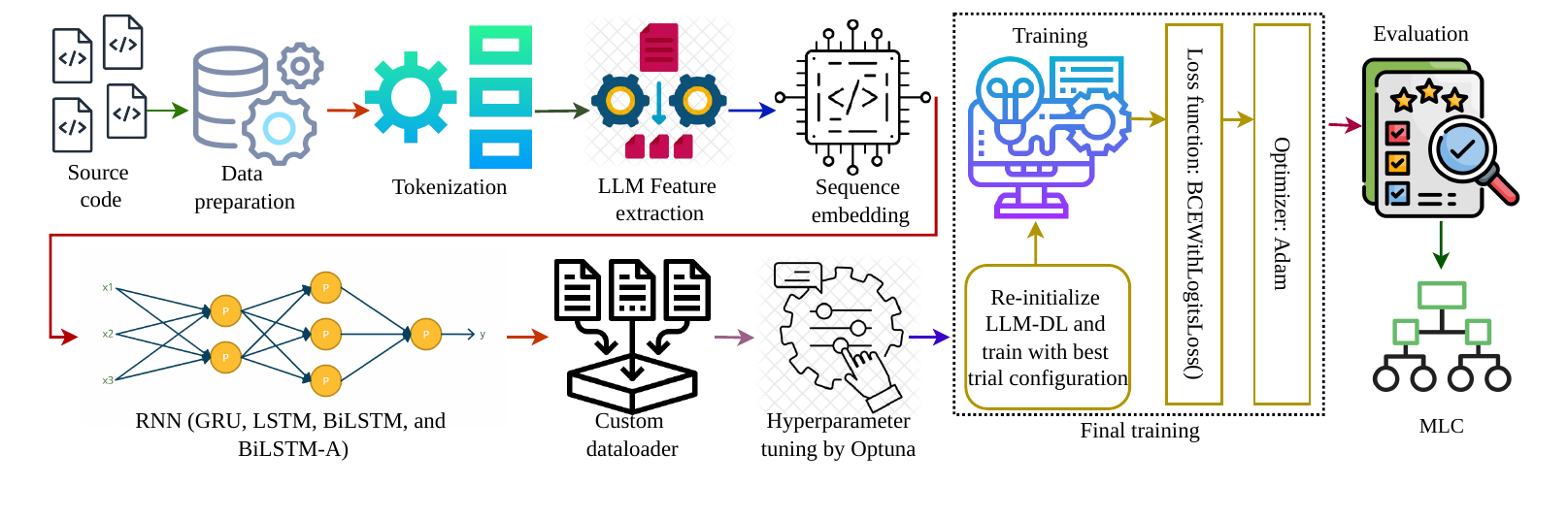}
  \caption{LLM-DL pipeline for MLEC}
  \label{PA}
\end{figure*}

This section describes how LLMs and DL architectures are combined for the MLEC task. The architecture is illustrated in~\figboxref{PA}, and the training and evaluation pipeline is described in~\algboxref{alg}.

\begin{algorithm}[h!]
\caption{MLEC using LLM-DL combinations}
\label{alg}
\begin{algorithmic}[1]
\State \textbf{Input:} Dataset $\mathcal{D} = \{(c_i, \mathbf{y}_i)\}_{i=1}^{N}$
\State \textbf{Output:} Trained model $f: c \mapsto \hat{\mathbf{y}}$ and evaluation metrics

\State \textbf{Step 1: Data Preparation}
\Statex \hspace{1em} -- \parbox[t]{\dimexpr\linewidth-2em}{Clean code snippets in $\mathcal{D}$, split into $\mathcal{D}_{\text{train}}$, $\mathcal{D}_{\text{val}}$, and $\mathcal{D}_{\text{test}}$} 

\State \textbf{Step 2: Tokenization}
\Statex \hspace{1em} -- \parbox[t]{\dimexpr\linewidth-2em}{For each $c_i \in \mathcal{D}$ do tokenize $c_i$ using LLM tokenizer:}
\Statex \hspace{2em} -- \parbox[t]{\dimexpr\linewidth-3em}{$\texttt{input\_ids}_i \;\in\; \mathbb{N}^{T}$, \quad $\texttt{attention\_mask}_i \;\in\; \{0,1\}^T$}
\Statex \hspace{1em} -- \parbox[t]{\dimexpr\linewidth-2em}{end for}

\State \textbf{Step 3: LLM Feature Extraction}
\Statex \hspace{1em} -- \parbox[t]{\dimexpr\linewidth-2em}{For each batch $(\mathbf{x}, \mathbf{m})$ from DataLoader do:}
\Statex \hspace{2em} -- \parbox[t]{\dimexpr\linewidth-3em}{Compute hidden representations using LLM:}
\Statex \hspace{3em} -- \parbox[t]{\dimexpr\linewidth-2em}{$\mathbf{H} = \text{LLM}(\mathbf{x}, \mathbf{m}) \in \mathbb{R}^{B \times T \times d}$}
\Statex \hspace{1em} -- \parbox[t]{\dimexpr\linewidth-2em}{end for}

\State \textbf{Step 4: Deep Learning Classifier}
\Statex \hspace{1em} -- \parbox[t]{\dimexpr\linewidth-2em}{Pass $\mathbf{H}$ into RNN module}
\Statex \hspace{1em} -- \parbox[t]{\dimexpr\linewidth-2em}{Apply: RNN $\rightarrow$ Max Pooling/Attention $\rightarrow$ Dropout $\rightarrow$ Dense}
\Statex \hspace{1em} -- \parbox[t]{\dimexpr\linewidth-2em}{Compute logits:\\
\hspace*{1em}{$\mathbf{z} = \mathbf{W}\mathbf{r} + \mathbf{b} \in \mathbb{R}^{B \times L}$}} 

\State \textbf{Step 5: Hyperparameter Tuning (Optuna)}
\Statex \hspace{1em} -- \parbox[t]{\dimexpr\linewidth-2em}{Repeat for $t = 1$ to $N_{\text{trial}}$ (e.g., $N_{\text{trial}}=10$) do:}
\Statex \hspace{2em} -- \parbox[t]{\dimexpr\linewidth-3em}{Sample trial parameters: \\ \hspace*{1em}$\theta_t = \{\eta, d_h, n_L, p_{\text{drop}}, b_s, d_{\text{bi}}\}$}
\Statex \hspace{2em} -- \parbox[t]{\dimexpr\linewidth-2em}{Train $f_{\theta_t}$ on $\mathcal{D}_{\text{train}}$ for $E_{\text{tune}}$ epochs}
\Statex \hspace{2em} -- \parbox[t]{\dimexpr\linewidth-3em}{Evaluate on $\mathcal{D}_{\text{val}}$ and report $F1_{\psi}^{(t)}$ to Optuna}
\Statex \hspace{1em} -- \parbox[t]{\dimexpr\linewidth-2em}{end for}
\Statex \hspace{1em} -- \parbox[t]{\dimexpr\linewidth-2em}{Select best: $\theta^* = \arg\max_t F1_{\psi}^{(t)}$}

\State \textbf{Step 6: Final Model Training}
\Statex \hspace{1em} -- \parbox[t]{\dimexpr\linewidth-2em}{Initialize model $f_{\theta^*}$ with best hyperparameters}
\Statex \hspace{1em} -- \parbox[t]{\dimexpr\linewidth-2em}{Train $f_{\theta^*}$ on $\mathcal{D}_{\text{train}}$ for $E$ epochs using Adam}
\Statex \hspace{1em} -- \parbox[t]{\dimexpr\linewidth-2em}{Loss function: $\mathcal{L} = \text{BCEWithLogitsLoss}(\mathbf{z}, \mathbf{y})$}

\State \textbf{Step 7: Evaluation}
\Statex \hspace{1em} -- \parbox[t]{\dimexpr\linewidth-2em}{Evaluate the trained model on $\mathcal{D}_{\text{test}}$}
\Statex \hspace{1em} -- \parbox[t]{\dimexpr\linewidth-2em}{Threshold predictions: $\hat{\mathbf{y}} = \sigma(\mathbf{z})$, $\tilde{\mathbf{y}} = \mathbb{I}[\hat{\mathbf{y}} > 0.5]$}

\end{algorithmic}
\end{algorithm}
\textbf{Step 1: Data Preparation}
Each element of the label vector $\mathbf{y}_i$ indicates the presence or absence of a corresponding error label under the multi-hot encoding used to construct the label matrix. Preprocessing removes unused imports, comments, redundant whitespace, and irrelevant characters while preserving case sensitivity and syntactic structure. The dataset is divided into a development subset $\mathcal{D}_{\text{dev}}$ and a test subset $\mathcal{D}_{\text{test}}$ using an 80:20 split, and 10\% of $\mathcal{D}_{\text{dev}}$ is held out for validation. The effective split is therefore 72\% training, 8\% validation, and 20\% testing. Validation data drive Optuna-based tuning and model selection; test data are used only for final evaluation.

\textbf{Step 2: Tokenization}
Each code sample $c_i$ is tokenized using the pre-trained tokenizer of the selected LLM, which maps $c_i$ to a fixed-length vector of token IDs and an attention mask:
\vspace{-0.35em}
\begin{equation}
\begin{aligned}
    \texttt{input\_ids}_i \in \mathbb{N}^T, \quad \\ \texttt{attention\_mask}_i \in \{0,1\}^T
\end{aligned}
\end{equation}

where $T$ is the maximum sequence length. Padding and truncation give uniform input
dimensions. Tensors are produced with \texttt{return\_tensors="pt"}, and the labels
$\mathbf{y}_i$ are cast to $\texttt{FloatTensor}$.

\textbf{Step 3: LLM Feature Extraction}
Tokenized inputs are passed to the LLM encoder $f_{\text{LLM}}$. Given a batch
$(\mathbf{X}, \mathbf{M})$, where $\mathbf{X} \in \mathbb{N}^{B \times T}$ holds the input IDs and $\mathbf{M} \in \{0,1\}^{B \times T}$ the attention masks, the encoder computes contextual representations:
\vspace{-0.35em}
\begin{equation}
    \mathbf{H} = f_{\text{LLM}}(\mathbf{X}, \mathbf{M}) \in \mathbb{R}^{B \times T \times d}
\end{equation}

where $B$ is the batch size and $d$ is the hidden dimensionality of the LLM. These
representations encode token-level semantics and inter-token dependencies. The full
sequence of embeddings is retained and passed to the downstream module rather than
reduced to a single pooled vector at this stage.

\textbf{Step 4: DL Classifier}
The contextual embeddings $\mathbf{H} \in \mathbb{R}^{B \times T \times d}$ are fed
into a recurrent sequence model $f_{\text{DL}}$ that models dependencies among token embeddings. Let $\mathbf{H}_t \in \mathbb{R}^{d}$ denote the embedding of the $t$-th token. The recurrent model produces hidden states:
\vspace{-0.35em}
\begin{equation}
    \mathbf{h}_t = f_{\text{RNN}}(\mathbf{h}_{t-1}, \mathbf{H}_t; \theta_{\text{RNN}}), \quad \text{for } t = 1, \dots, T
\end{equation}

where $\theta_{\text{RNN}}$ are the trainable recurrent parameters. Under a bidirectional setting, forward and backward hidden states are concatenated:
\begin{equation}
    \mathbf{h}_t = [\overrightarrow{\mathbf{h}}_t \, \| \, \overleftarrow{\mathbf{h}}_t] \in \mathbb{R}^{2d_h}
\end{equation}

where $d_h$ is the hidden size of each unidirectional layer. The hidden-state
sequence is aggregated into a sequence-level representation $\mathbf{r}$. For GRU,
LSTM, and BiLSTM variants, max pooling is applied over the sequence dimension:
\vspace{-0.4em}
\begin{equation}
    \mathbf{r} = \max_{1 \leq t \leq T} \mathbf{h}_t
    \label{lab1}
\end{equation}

For BiLSTM-A, $\mathbf{r}$ is the attention-weighted context vector produced by the
additive attention module. The representation $\mathbf{r}$ passes through a dropout
layer and a fully connected layer to produce logits:
\vspace{-0.5em}
\begin{equation}
    \mathbf{z} = \mathbf{W}\mathbf{r} + \mathbf{b} \in \mathbb{R}^{B \times L}
    \label{lab2}
\end{equation}

where $\mathbf{W} \in \mathbb{R}^{L \times d_r}$ and $\mathbf{b} \in \mathbb{R}^{L}$ are learnable parameters, $d_r$ is the dimensionality of $\mathbf{r}$, and $L$ is the number of labels. Training uses \texttt{BCEWithLogitsLoss}, which operates directly on logits.

\textbf{Step 5: Hyperparameter Tuning (Optuna)}
Optuna-based tuning identifies the best configuration for each LLM–DL combination. In each trial $t$, a configuration is sampled from a predefined search space:
\vspace{-0.4em}
\begin{equation}
\begin{aligned}
    \theta_t = \{\eta, d_h, n_L, p_{\text{drop}}, b_s, d_{\text{bi}}\}
\end{aligned}
\end{equation}

where $\eta$ is the learning rate, $d_h$ is the hidden dimension, $n_L\in \{1, 2\}$ is the number of recurrent layers, $p_{\text{drop}}$ is the dropout probability, $b_s$ is the batch size, and $d_{\text{bi}}$ denotes the bidirectional setting.\footnote{The \texttt{bidirectional} flag applies to GRU and LSTM architectures; BiLSTM and BiLSTM-A are inherently bidirectional.} Each sampled configuration $f_{\theta_t}$ is trained on $\mathcal{D}_{\text{train}}$ for $E_{\text{tune}} = 1$ epoch and evaluated on $\mathcal{D}_{\text{val}}$. Validation performance is measured by the weighted F1-score, $F1_{\psi}^{(t)}$, and reported to Optuna. The best configuration is then selected as:
\vspace{-0.5mm}
\begin{equation}
    \theta^* = \arg\max_{t \in \{1,\ldots,N_{\text{trial}}\}} F1_{\psi}^{(t)}
\end{equation}

where $N_{\text{trial}} = 10$ in this study.

\textbf{Step 6: Final Model Training}
The model $f_{\theta^*}$ is re-initialized with the selected configuration $\theta^*$ and trained on $\mathcal{D}_{\text{train}}$ for $E = 20$ epochs using the Adam optimizer, at the tuned learning rate $\eta$ and the batch size listed in \tabboxref{BHP}. The model outputs logits $\mathbf{z}$, and the training objective is optimized using \texttt{BCEWithLogitsLoss}. For clarity, this loss can be expressed using the sigmoid-transformed probability $\hat{y}_{ij} = \sigma(z_{ij})$ as:
\vspace{-0.5em}
\begin{equation}
    \mathcal{L} = -\frac{1}{B \cdot L} \sum_{i=1}^{B} \sum_{j=1}^{L}
    \left[
    y_{ij}\log(\hat{y}_{ij}) + (1-y_{ij})\log(1-\hat{y}_{ij})
    \right]
\end{equation}

where $B$ is the batch size, $L$ is the number of labels, $y_{ij}$ is the ground-truth binary label, and $\hat{y}_{ij}$ is the predicted probability after applying sigmoid to the logit $z_{ij}$. In implementation, \texttt{BCEWithLogitsLoss} receives logits directly and applies the sigmoid internally in a numerically stable form. Training and validation performance are monitored throughout.

\textbf{Step 7: Evaluation}
The trained model is evaluated on $\mathcal{D}_{\text{test}}$. Logits $\mathbf{z}$
are converted to probabilities:
\vspace{-0.5em}
\begin{equation}
    \hat{\mathbf{y}} = \sigma(\mathbf{z})
\end{equation}

and thresholded at $\tau = 0.5$ to obtain binary multi-label predictions:
\vspace{-0.5em}
\begin{equation}
    \tilde{\mathbf{y}} = \mathbb{I}[\hat{\mathbf{y}} > \tau]
\end{equation}

Test-set performance is then reported using the metrics defined in Section~\ref{Experiment}.

\subsection{Baseline and Ablation Architectures}
\label{Ablation Architectures}
Section~\ref{Results} compares the LLM-DL combinations described above against the baseline and ablation configurations defined below, all of which share the same encoder, tokenization, loss function, optimizer, data split, and training schedule, and differ only in what follows the encoder. Each configuration receives its own Optuna search under the same budget used for the LLM–DL combinations, so no configuration is evaluated with another's hyperparameters.

\textbf{Linear classification head}
The recurrent decoder is removed. The encoder output $\mathbf{H}$ is aggregated
directly over the token dimension by an element-wise maximum, and the pooled vector
passes through dropout and a linear projection of the same form as~\eqref{lab2}:
\vspace{-0.35em}
\begin{equation}
\mathbf{r} = \max_{1 \leq t \leq T} \mathbf{H}_t, \qquad
\mathbf{z} = \mathbf{W}_h \, \text{dropout}(\mathbf{r}) + \mathbf{b}_h
\label{lab3}
\end{equation}

where $\mathbf{W}_h \in \mathbb{R}^{L \times d}$ and $\mathbf{b}_h \in \mathbb{R}^{L}$
are learnable parameters. Because the encoder and every other stage of the pipeline are unchanged, the comparison between this configuration and its LLM–DL counterpart measures the contribution of the recurrent decoder.

\textbf{Pooling variants}
To examine how the aggregation step affects the head, the maximum in~\eqref{lab3} is replaced by two alternatives. Mean pooling averages over non-padding tokens, and additive attention pooling learns a scalar score for each token and returns the weighted sum:
\vspace{-0.35em}
\begin{equation}
\begin{aligned}
    \mathbf{r}_{\text{mean}} &= \frac{\sum_{t=1}^{T} m_t \mathbf{H}_t}{\sum_{t=1}^{T} m_t}, \qquad
    s_t = \mathbf{w}_2^{\top} \tanh(\mathbf{W}_1 \mathbf{H}_t), \\
    \alpha_t &= \frac{\exp(s_t)}{\sum_{t'=1}^{T} \exp(s_{t'})}, \qquad
    \mathbf{r}_{\text{attn}} = \sum_{t=1}^{T} \alpha_t \mathbf{H}_t
    \label{lab4}
\end{aligned}
\end{equation}

where $m_t \in \{0,1\}$ denotes the attention mask, and $\mathbf{W}_1 \in
\mathbb{R}^{d_a \times d}$ and $\mathbf{w}_2 \in \mathbb{R}^{d_a}$ are learnable with
$d_a = 128$. Padding positions are excluded from the softmax.

\textbf{Label-attention head}
Equations~\eqref{lab3} and~\eqref{lab4} predict every label from a single shared representation, so dependence among labels is captured only implicitly through the encoder. The label-attention head retains the max-pooled representation of~\eqref{lab3} and replaces the final linear layer with a block in which the labels interact directly. The pooled vector is projected into $L$ label-specific embeddings that attend to one another:
\vspace{-0.35em}
\begin{equation}
\begin{aligned}
    \mathbf{E} &= \text{reshape}_{L \times d_l}\!\left(\mathbf{W}_p\,\text{dropout}(\mathbf{r}) + \mathbf{b}_p\right), \\
    \tilde{\mathbf{E}} &= \text{LayerNorm}\!\left(\mathbf{E} + \text{dropout}\!\left(\text{MHA}(\mathbf{E}, \mathbf{E}, \mathbf{E})\right)\right), \\
    z_l &= \mathbf{w}_c^{\top} \tilde{\mathbf{E}}_l + b_c, \qquad l = 1, \dots, L
    \label{lab5}
\end{aligned}
\end{equation}

where $\mathbf{r}$ is the max-pooled encoder output from~\eqref{lab3}, $\mathbf{W}_p \in \mathbb{R}^{(L \cdot d_l) \times d}$ projects it into $L$ embeddings of size $d_l$, and $\tilde{\mathbf{E}}_l \in \mathbb{R}^{d_l}$ is the representation of label $l$ after interaction. MHA denotes multi-head self-attention applied over the label dimension with $n_h$ heads, whose $L \times L$ attention weights represent the learned dependencies among error types, and $\mathbf{w}_c \in \mathbb{R}^{d_l}$ with $b_c \in \mathbb{R}$ form a classifier shared across labels. The pair $(d_l, n_h)$ is tuned jointly over $\{(64,4), (64,8), (96,4), (128,4), (128,8)\}$ together with the learning rate, dropout rate, and batch size.

\section{Experiments}
\label{Experiment}
\subsection{Implementation Settings}
The experiments were conducted on a 64-bit Windows 11 Pro operating system. The hardware configuration comprised an AMD Ryzen Threadripper 2970WX 24-core processor (3.00 GHz), 64 GB RAM, an NVIDIA GeForce RTX 2080 Ti GPU with 12 GB VRAM, and 1 TB of storage. The 32 LLM-DL combinations in Section~\ref{Results} are each trained once, as in the original evaluation protocol. The baseline and ablation experiments that follow
are trained across multiple random seeds, five for the CodeT5+ configurations and three for the supporting encoders, using the seed values 13, 42, 123, 2024, and 31337. For these experiments, results are reported as mean and standard deviation, and paired comparisons across matched seeds are assessed with a two-tailed paired $t$-test. Optuna tuning for the seeded experiments uses a fixed sampler seed, so each search is reproducible.

\subsection{Dataset and Preprocessing}
The error-labeled dataset introduced in~\cite{amin2025source} is utilized in this study. It consists of 95,631 pairs of erroneous and accepted code samples, collected from 44 introductory programming problems in the ``Introduction to Programming 1 (ITP1)'' course on the Aizu Online Judge (AOJ) system~\cite{watanobe2018aizu, aizu_developers}. Each pair couples a student-submitted erroneous solution with a corresponding accepted solution. Under the original 55-label scheme, each pair contains 3.47 errors on average (\(\pm\)2.69), with at least one error per pair. As the erroneous code was sourced from AOJ, all submissions were automatically evaluated against test input/output cases, and the corresponding error verdicts were recorded. Basic statistics of the source dataset are shown in~\figboxref{DI}\footnote{The Avg. char-based similarity and Avg. token-based similarity values are expressed as percentages, and bar label values are presented in $\pm SD$.}

\begin{figure}[t]
\hspace{-3mm}
     \centering
     \captionsetup{justification=centering}
         \begin{tikzpicture}[scale=1]
            \input{Graphs_Figures/DI}
        \end{tikzpicture}
        \caption{Basic statistics of the source error label dataset}
         \label{DI}
    \end{figure}

Two preparation steps change these counts before modeling. First, the 55 original labels are summarized into 11 labels by frequency and semantic similarity, which
merges errors that share a summarized category. Second, the preprocessing steps in
\tabboxref{DP} remove duplicate and invalid rows. The resulting dataset contains 82,774 samples carrying 217,363 label occurrences, an average of 2.63 labels per sample ($\pm$1.28), with a minimum of one and a maximum of nine. Label density, the mean fraction of active labels per sample, is 0.239. All experiments in this study use these 82,774 samples, and the 80:20 split yields 16,555 test samples. Further detail on the original and summarized error label sets is available in~\cite{amin2024multi}. \figboxref{EF} shows the frequency of the 11 summarized labels across the preprocessed samples. Support is highly uneven: the most frequent label, \texttt{input\_output}, occurs in 45,888 samples, whereas the least frequent, \texttt{function\_definition}, occurs in 454, a ratio of roughly 101 to 1. This imbalance is the principal reason macro-averaged and weighted-averaged results differ throughout Section~\ref{Results}.

\begin{figure}[h]
\hspace{-3mm}
     \centering
     \captionsetup{justification=centering}
         \begin{tikzpicture}[scale=1]
            \input{Graphs_Figures/EF}
        \end{tikzpicture}
        \caption{Frequency of the 11 summarized error labels}
         \label{EF}
    \end{figure}

\begin{table}[t]
\caption{Data preparation and processing steps}
\setlength{\arrayrulewidth}{0.2mm}
\setlength{\tabcolsep}{1.9pt}
\renewcommand{\arraystretch}{1}
\label{DP}
\centering
\begin{tabular}{
>{\centering\arraybackslash}m{1.45cm}||
>{\centering\arraybackslash}m{4.25cm}||
>{\centering\arraybackslash}m{2.4cm}
}
\hline \hline
\textbf{Step} & \textbf{Description} & \textbf{Output}  \\ 
\hline \hline

Data Loading 
& Load dataset containing code samples and error labels 
& Raw dataset for preprocessing \\ 
\hline

Data Cleaning 
& Normalize text by removing extra white lines, comments, and unused imports, preserving syntax/case sensitivity and indentation 
& \shortstack{Cleaned dataset, \\optimized for \\model encoding} \\ 
\hline

Handling Missing Data 
& \shortstack{Remove empty or invalid rows \\and verify label-vector consistency} 
& \shortstack{Dataset with no \\missing/invalid \\entries} \\ 
\hline

Tokenization 
& \shortstack{Tokenize code using corresponding \\LLM tokenizer. Output generated \\by using \texttt{return\_tensors=pt}} 
& \texttt{input\_ids} and \texttt{attention\_mask} tensors \\ 
\hline

Label Encoding 
& Convert labels into a binary indicator matrix using multi-hot encoding scheme 
& \shortstack{Binary label \\vectors (1 if present \\and 0 otherwise)} \\ 
\hline

Dataset Splitting 
& \shortstack{Split into development (80\%) \\and test (20\%); use 10\% of the \\development set for validation} 
& Final subsets: train/validation/test (72\%/8\%/20\%) \\ 
\hline

Creating DataLoader 
& \shortstack{Wrap tokenized data and labels \\into a custom PyTorch dataset} 
& \shortstack{Batches of token\\-label tensors for \\train/validation/test} \\ 
\hline \hline
\end{tabular}
\end{table}

\begin{figure}[t!]
  \centering
  \includegraphics[width=1\linewidth]{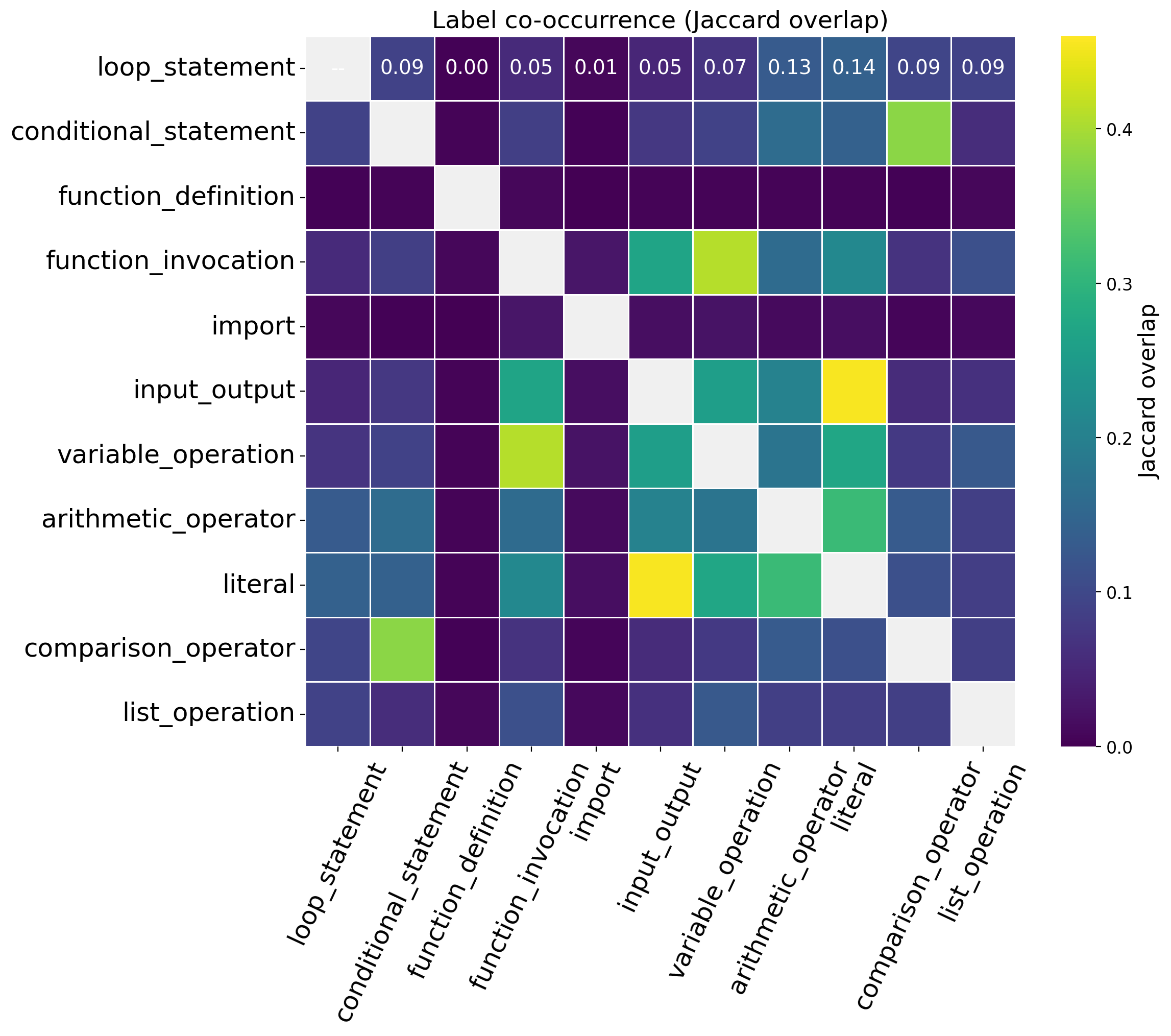}
  \caption{Pairwise label co-occurrence measured by Jaccard overlap}
  \label{CO}
\end{figure}

Beyond individual frequencies, the labels co-occur in semantically coherent ways,
and this dependence is concentrated rather than uniform. Measured by the pairwise Jaccard coefficient, three pairs stand clearly above the rest: \texttt{input\_output} with \texttt{literal} (0.455), \texttt{function\_invocation} with \texttt{variable\_operation} (0.410), and \texttt{conditional\_statement} with \texttt{comparison\_operator} (0.380). The two rarest labels are close to independent of every other label, reaching at most 0.026 for \texttt{import} and 0.009 for \texttt{function\_definition}. As~\figboxref{CO} shows, high overlap is restricted to a handful of frequent-label pairs, while the rows for \texttt{import} and \texttt{function\_definition} remain low against every other label. The diagonal is omitted, since self-overlap is always one.

\subsection{Evaluation Metrics}
To evaluate model performance across different aspects of prediction quality, the following metric suite is used~\cite{si2023multi, amin2025source, lyu2024vsm}. Let \( \mathbf{y}_i = [y_{i1}, y_{i2}, \dots, y_{iL}] \in \{0,1\}^L \) represent the ground-truth label vector for the \( i \)-th instance, where \( L \) is the total number of labels. The model outputs logits \( \mathbf{z}_i \), which are converted into probability scores \( \hat{\mathbf{p}}_i = \sigma(\mathbf{z}_i) \in [0,1]^L \). Binary predictions are then obtained as \( \tilde{\mathbf{y}}_i = \mathbb{I}[\hat{\mathbf{p}}_i > 0.5] \). All evaluation metrics are computed on the test set containing \( N \) instances.

\subsubsection{Average Accuracy ($AvgAcc$)}
$AvgAcc$ reflects the mean proportion of correct label-wise predictions per sample, including both TP and TN.
\vspace{-0.7em}
\begin{equation}
\text{AvgAcc} = \frac{1}{N} \sum_{i=1}^{N} \left( \frac{1}{L} \sum_{j=1}^{L} \mathbb{I}(y_{ij} = \tilde{y}_{ij}) \right)
\end{equation}
where \( \mathbb{I}(y_{ij} = \tilde{y}_{ij}) \) returns 1 when the predicted and ground-truth labels match. Higher values indicate better label-wise correctness.

\subsubsection{Exact Match Accuracy ($EMAcc$)}
$EMAcc$ calculates the proportion of instances whose predicted label set exactly matches the true label set.
\vspace{-0.5em}
\begin{equation}
\text{EMAcc} = \frac{1}{N} \sum_{i=1}^{N} \mathbb{I}(\tilde{\mathbf{y}}_i = \mathbf{y}_i)
\end{equation}
where \( \mathbb{I}(\cdot) \) returns 1 if the full predicted label vector is identical to the ground truth, and 0 otherwise. Higher values indicate stricter multi-label prediction correctness.

\subsubsection{One Error ($OE$)}
$OE$ evaluates whether the top-ranked predicted label is not among the true labels. It measures the reliability of the model's most confident prediction.
\begin{equation}
\text{OE} = \frac{1}{N} \sum_{i=1}^{N} \mathbb{I}\left(\arg\max_j \hat{p}_{ij} \notin \{j \mid y_{ij} = 1\}\right)
\end{equation}
where \( \hat{p}_{ij} \) denotes the predicted probability for label \( j \) of instance \( i \) and \(\arg\max_j \hat{p}_{ij}\) selects the top-scored label. This metric ranges from 0 to 1, with lower values indicating better performance.

\subsubsection{Precision ($P$), Recall ($R$), and F1 Score ($F1$)}
These metrics assess the quality of predictions at the label level. Two averaging strategies are used: \textbf{macro-average} ($\bar{\mu}$), which treats all labels equally, and \textbf{weighted-average} ($\psi$), which weights each label according to its support.

\subsubsection*{Macro Averaged Metrics}
\begin{align}
P_{\bar{\mu}} &= \frac{1}{L} \sum_{j=1}^{L} \frac{TP_j}{TP_j + FP_j} \\
R_{\bar{\mu}} &= \frac{1}{L} \sum_{j=1}^{L} \frac{TP_j}{TP_j + FN_j} \\ 
F1_{\bar{\mu}} &= \frac{1}{L} \sum_{j=1}^{L} \frac{2 \cdot TP_j}{2 \cdot TP_j + FP_j + FN_j}
\end{align}

\subsubsection*{Weighted Averaged Metrics}
\begin{align}
P_{\psi} &= \sum_{j=1}^{L} w_j \cdot \frac{TP_j}{TP_j + FP_j} \\
R_{\psi} &= \sum_{j=1}^{L} w_j \cdot \frac{TP_j}{TP_j + FN_j} \\
F1_{\psi} &= \sum_{j=1}^{L} w_j \cdot \frac{2 \cdot TP_j}{2 \cdot TP_j + FP_j + FN_j}
\end{align}
where \( TP_j \), \( FP_j \), and \( FN_j \) denote the true positives, false positives, and false negatives for label \( j \), respectively, and \( w_j = \frac{\sum_{i=1}^N y_{ij}}{\sum_{j=1}^L \sum_{i=1}^N y_{ij}} \) represents frequency-based weight for label \( j \). $F1$ is the harmonic mean of $P$ and $R$ (higher is better), and weighted metrics are useful when label frequencies are imbalanced.

\subsubsection{Hamming Loss ($HM$)}
$HM$ quantifies the fraction of labels that are incorrectly predicted across all instances and labels.
\begin{equation}
\text{HM} = \frac{1}{N \cdot L} \sum_{i=1}^{N} \sum_{j=1}^{L} \mathbb{I}(\tilde{y}_{ij} \neq y_{ij})
\end{equation}
Lower values are desirable, with zero indicating perfect label-wise predictions.

\subsubsection{Jaccard Similarity Score ($J_s$)}
$J_s$ is a widely used metric in MLC that measures the overlap between the predicted and ground-truth positive label sets. It is defined as the size of the intersection divided by the size of the union of predicted and actual positive labels, computed per sample and averaged over the dataset.
\vspace{-0.6em}
\begin{equation}
\text{J}_{\text{s}} = \frac{1}{N} \sum_{i=1}^{N} \frac{|\tilde{\mathbf{y}}_i \cap \mathbf{y}_i|}{|\tilde{\mathbf{y}}_i \cup \mathbf{y}_i|}
\end{equation}
where \( \tilde{\mathbf{y}}_i \) denotes the predicted positive label set and \( \mathbf{y}_i \) denotes the ground-truth positive label set for instance \( i \). Unlike $AvgAcc$, $J_s$ considers only the positive labels. It ignores TN and penalizes both FP and FN, making it more robust in imbalanced settings.

\subsubsection{ROC-AUC Score}
The Area Under the Receiver Operating Characteristic Curve (ROC-AUC) evaluates the model's ability to distinguish positive and negative cases using probability scores \( \hat{p}_{ij} \). Three variants are considered.

\textbf{Micro-averaged} aggregates predictions across all labels:
\vspace{-0.2em}
\begin{equation}
    \text{ROC-AUC}_{\mu} = \text{AUC}\left(\text{Flatten}(y_{ij}), \text{Flatten}(\hat{p}_{ij})\right)
\end{equation}

\textbf{Macro-averaged} computes AUC independently for each label and then averages across labels:
\vspace{-0.4em}
\begin{equation}
    \text{ROC-AUC}_{\bar{\mu}} = \frac{1}{L} \sum_{j=1}^{L} \text{AUC}(y_{\cdot j}, \hat{p}_{\cdot j})
\end{equation}

\textbf{Weighted-averaged} weights each label's AUC by its support:
\vspace{-0.8em}
\begin{equation}
    \text{ROC-AUC}_{\psi} = \sum_{j=1}^{L} w_j \cdot \text{AUC}(y_{\cdot j}, \hat{p}_{\cdot j})
\end{equation}

Values range from 0.0 to 1.0, with higher scores indicating stronger discrimination between positive and negative cases.

\subsection{Hyperparameter}
Hyperparameter selection stabilizes the learning behavior of the LLM-DL models, especially when modeling source code with complex token, function, variable, and operator structures. Optuna-based tuning is performed for each LLM-DL combination to identify effective settings for the decoder and classifier components. The general search space and fixed settings used for tuning and final training are summarized in~\tabboxref{HP} \footnote{The $lr$ range was adjusted only for RoBERTa\_LR experiments.}. The best-performing hyperparameter configurations for each architecture are listed in~\tabboxref{BHP}. For each model combination, the best trial ($id$) is selected based on the validation weighted F1-score, denoted as $F1_{\psi}$ ($ts$). The tuned parameters include hidden dimension ($hd$), number of recurrent layers ($\# L$), dropout rate ($dr$), learning rate ($lr$), batch size ($bs$), and bidirectional setting ($bd$). These parameters are optimized to improve MLC performance while reducing overfitting across different LLM-DL architectures.

\begin{table}[h]
\caption{Hyperparameter settings for the experiment}
\setlength{\arrayrulewidth}{0.2mm}
\setlength{\tabcolsep}{3.2pt}
\renewcommand{\arraystretch}{1}
\label{HP}
\begin{tabular}{c||c}
\hline \hline
\textbf{Hyperparameter}          & \textbf{Search Space / Value}    \\ \hline \hline
Tokenizer               & Model-specific tokenizer        \\ \hline
Output Layer            & Linear ($d_r$, $L$), where $L=11$        \\ \hline
Hidden Dimension        & \{64, 128, 256\}        \\ \hline
Number of Layers        & \{1, 2\} ; Dropout rate: [0.1, 0.3] (Uniform)   \\ \hline
Maximum Sequence Length & 256      \\ \hline
Batch Size        & \{4, 8, 16\}    \\ \hline
Bidirectional           & \{True, False\}      \\ \hline
Optimizer               & Adam ; Loss: BCEWithLogitsLoss                  \\ \hline
Activation Function     & \shortstack{Sigmoid for probability conversion \\during evaluation}                 \\ \hline
Learning Rate           & [1e-5, 1e-1] (LogUniform)             \\ \hline
Number of Epochs        & 20 for final training                  \\ \hline \hline
\end{tabular}
\end{table}

\begin{table}[]
\caption{The best hyperparameters of the models}
\setlength{\arrayrulewidth}{0.2mm}
\setlength{\tabcolsep}{2.8pt}
\renewcommand{\arraystretch}{1}
\label{BHP}
\begin{tabular}{cc||c||c||cccccc}
\hline
\multicolumn{2}{c||}{\multirow{2}{*}{\textbf{{\shortstack{Model\\Combination}}}}}        & \multirow{2}{*}{\textbf{$id$}} & \multirow{2}{*}{\textbf{$ts$}} & \multicolumn{6}{c}{\textbf{Best Hyperparameters}}      \\ \cline{5-10} 
\multicolumn{2}{c||}{}                                          &                        &                           & \multicolumn{1}{c|}{\textbf{$hd$}}  & \multicolumn{1}{c|}{\textbf{$\# L$}} & \multicolumn{1}{c|}{\textbf{$dr$}}  & \multicolumn{1}{c|}{\textbf{$lr$}}             & \multicolumn{1}{c|}{\textbf{$bs$}} & \multicolumn{1}{c}{\textbf{$bd$}} \\ \hline \hline
\multicolumn{1}{c|}{\multirow{4}{*}{\shortstack{Code\\T5}}}        & GRU      & 8                      & 0.7153                   & \multicolumn{1}{c|}{128} & \multicolumn{1}{c|}{2}    & \multicolumn{1}{c|}{0.1260} & \multicolumn{1}{c|}{$9.02e^{-5}$}  & \multicolumn{1}{c|}{8} & F \\ 
\multicolumn{1}{c|}{}                               & LSTM     & 2                      & 0.7365                   & \multicolumn{1}{c|}{256} & \multicolumn{1}{c|}{1}    & \multicolumn{1}{c|}{0.1827} & \multicolumn{1}{c|}{$3.67e^{-5}$}  & \multicolumn{1}{c|}{8} & T  \\ 
\multicolumn{1}{c|}{}                               & BiLSTM   & 2                      & 0.7239                   & \multicolumn{1}{c|}{256} & \multicolumn{1}{c|}{1}    & \multicolumn{1}{c|}{0.1134} & \multicolumn{1}{c|}{$2.71e^{-5}$} & \multicolumn{1}{c|}{4} & T  \\ 
\multicolumn{1}{c|}{}                               & BiLSTM-A & 7                      & 0.6755                   & \multicolumn{1}{c|}{256} & \multicolumn{1}{c|}{1}    & \multicolumn{1}{c|}{0.1138} & \multicolumn{1}{c|}{$6.06e^{-4}$}  & \multicolumn{1}{c|}{8} & T  \\ \hline
\multicolumn{1}{c|}{\multirow{4}{*}{\shortstack{Graph\\Code\\BERT}}} & GRU      & 6                      & 0.7272                   & \multicolumn{1}{c|}{128} & \multicolumn{1}{c|}{1}    & \multicolumn{1}{c|}{0.1304} & \multicolumn{1}{c|}{$2.51e^{-5}$} & \multicolumn{1}{c|}{8} & F \\ 
\multicolumn{1}{c|}{}                               & LSTM     & 2                      & 0.3802                   & \multicolumn{1}{c|}{128} & \multicolumn{1}{c|}{2}    & \multicolumn{1}{c|}{0.1846} & \multicolumn{1}{c|}{$2.43e^{-2}$}   & \multicolumn{1}{c|}{8} & F \\ 
\multicolumn{1}{c|}{}                               & BiLSTM   & 6                      & 0.6181                   & \multicolumn{1}{c|}{128} & \multicolumn{1}{c|}{2}    & \multicolumn{1}{c|}{0.2710} & \multicolumn{1}{c|}{$1.16e^{-5}$} & \multicolumn{1}{c|}{4} & T  \\ 
\multicolumn{1}{c|}{}                               & BiLSTM-A & 2                      & 0.6367                   & \multicolumn{1}{c|}{256} & \multicolumn{1}{c|}{2}    & \multicolumn{1}{c|}{0.1069} & \multicolumn{1}{c|}{$5.16e^{-4}$}  & \multicolumn{1}{c|}{8} & T  \\ \hline
\multicolumn{1}{c|}{\multirow{4}{*}{\shortstack{Code\\T5+}}}       & GRU      & 3                      & 0.7249                   & \multicolumn{1}{c|}{128} & \multicolumn{1}{c|}{2}    & \multicolumn{1}{c|}{0.1696} & \multicolumn{1}{c|}{$2.29e^{-5}$} & \multicolumn{1}{c|}{4} & T  \\ 
\multicolumn{1}{c|}{}                               & LSTM     & 4                      & 0.5026                   & \multicolumn{1}{c|}{128} & \multicolumn{1}{c|}{1}    & \multicolumn{1}{c|}{0.2203} & \multicolumn{1}{c|}{$2.66e^{-4}$}   & \multicolumn{1}{c|}{4} & T  \\ 
\multicolumn{1}{c|}{}                               & BiLSTM   & 1                      & 0.6601                   & \multicolumn{1}{c|}{128} & \multicolumn{1}{c|}{2}    & \multicolumn{1}{c|}{0.2578} & \multicolumn{1}{c|}{$2.98e^{-5}$} & \multicolumn{1}{c|}{8} & T  \\ 
\multicolumn{1}{c|}{}                               & BiLSTM-A & 9                      & 0.6382                   & \multicolumn{1}{c|}{256} & \multicolumn{1}{c|}{2}    & \multicolumn{1}{c|}{0.1601} & \multicolumn{1}{c|}{$4.89e^{-4}$}   & \multicolumn{1}{c|}{4} & T  \\ \hline
\multicolumn{1}{c|}{\multirow{4}{*}{\shortstack{Uni\\Xcoder}}}     & GRU      & 0                      & 0.6058                   & \multicolumn{1}{c|}{128} & \multicolumn{1}{c|}{2}    & \multicolumn{1}{c|}{0.1475} & \multicolumn{1}{c|}{$3.28e^{-4}$}   & \multicolumn{1}{c|}{4} & F \\ 
\multicolumn{1}{c|}{}                               & LSTM     & 6                      & 0.5798                   & \multicolumn{1}{c|}{128} & \multicolumn{1}{c|}{2}    & \multicolumn{1}{c|}{0.2334} & \multicolumn{1}{c|}{$2.29e^{-4}$} & \multicolumn{1}{c|}{4} & F \\ 
\multicolumn{1}{c|}{}                               & BiLSTM   & 6                      & 0.6098                   & \multicolumn{1}{c|}{128} & \multicolumn{1}{c|}{2}    & \multicolumn{1}{c|}{0.2747} & \multicolumn{1}{c|}{$6.82e^{-4}$}  & \multicolumn{1}{c|}{4} & T  \\ 
\multicolumn{1}{c|}{}                               & BiLSTM-A & 3                      & 0.5952                   & \multicolumn{1}{c|}{128} & \multicolumn{1}{c|}{2}    & \multicolumn{1}{c|}{0.1475} & \multicolumn{1}{c|}{$1.79e^{-3}$}   & \multicolumn{1}{c|}{16} & T  \\ \hline
\multicolumn{1}{c|}{\multirow{4}{*}{\shortstack{Ro\\BERTa}}}       & GRU      & 4                      & 0.6448                   & \multicolumn{1}{c|}{256} & \multicolumn{1}{c|}{2}    & \multicolumn{1}{c|}{0.2323} & \multicolumn{1}{c|}{$1.32e^{-5}$}  & \multicolumn{1}{c|}{4} & T  \\ 
\multicolumn{1}{c|}{}                               & LSTM     & 0                      & 0.3802                   & \multicolumn{1}{c|}{128} & \multicolumn{1}{c|}{2}    & \multicolumn{1}{c|}{0.1941} & \multicolumn{1}{c|}{$5.86e^{-2}$}   & \multicolumn{1}{c|}{16} & F \\ 
\multicolumn{1}{c|}{}                               & BiLSTM   & 7                      & 0.5295                   & \multicolumn{1}{c|}{128} & \multicolumn{1}{c|}{2}    & \multicolumn{1}{c|}{0.1844} & \multicolumn{1}{c|}{$3.03e^{-5}$}  & \multicolumn{1}{c|}{8} & T  \\ 
\multicolumn{1}{c|}{}                               & BiLSTM-A & 3                      & 0.2648                   & \multicolumn{1}{c|}{128} & \multicolumn{1}{c|}{2}    & \multicolumn{1}{c|}{0.1306} & \multicolumn{1}{c|}{$5.40e^{-3}$}   & \multicolumn{1}{c|}{4} & T  \\ \hline
\multicolumn{1}{c|}{\multirow{4}{*}{\shortstack{Ro\\BERTa\\\_LR}}}   & GRU      & 6                      & 0.6516                   & \multicolumn{1}{c|}{64}  & \multicolumn{1}{c|}{1}    & \multicolumn{1}{c|}{0.1475} & \multicolumn{1}{c|}{$1.52e^{-5}$} & \multicolumn{1}{c|}{4} & T  \\ 
\multicolumn{1}{c|}{}                               & LSTM     & 4                      & 0.6572                   & \multicolumn{1}{c|}{256} & \multicolumn{1}{c|}{1}    & \multicolumn{1}{c|}{0.2920} & \multicolumn{1}{c|}{$1.48e^{-5}$} & \multicolumn{1}{c|}{8} & F \\ 
\multicolumn{1}{c|}{}                               & BiLSTM   & 0                      & 0.6124                   & \multicolumn{1}{c|}{128} & \multicolumn{1}{c|}{1}    & \multicolumn{1}{c|}{0.2919} & \multicolumn{1}{c|}{$1.08e^{-5}$} & \multicolumn{1}{c|}{16} & T  \\ 
\multicolumn{1}{c|}{}                               & BiLSTM-A & 8                      & 0.5731                   & \multicolumn{1}{c|}{128} & \multicolumn{1}{c|}{2}    & \multicolumn{1}{c|}{0.1769} & \multicolumn{1}{c|}{$1.40e^{-5}$} & \multicolumn{1}{c|}{4} & T  \\ \hline
\multicolumn{1}{c|}{\multirow{4}{*}{\shortstack{PL\\BART}}}        & GRU      & 5                      & 0.6723                   & \multicolumn{1}{c|}{128} & \multicolumn{1}{c|}{1}    & \multicolumn{1}{c|}{0.1776} & \multicolumn{1}{c|}{$2.38e^{-5}$} & \multicolumn{1}{c|}{4} & T  \\ 
\multicolumn{1}{c|}{}                               & LSTM     & 2                      & 0.6098                   & \multicolumn{1}{c|}{256} & \multicolumn{1}{c|}{2}    & \multicolumn{1}{c|}{0.1057} & \multicolumn{1}{c|}{$1.25e^{-5}$} & \multicolumn{1}{c|}{4} & F \\ 
\multicolumn{1}{c|}{}                               & BiLSTM   & 8                      & 0.6738                   & \multicolumn{1}{c|}{256} & \multicolumn{1}{c|}{1}    & \multicolumn{1}{c|}{0.1411} & \multicolumn{1}{c|}{$4.28e^{-5}$}  & \multicolumn{1}{c|}{16} & T  \\ 
\multicolumn{1}{c|}{}                               & BiLSTM-A & 5                      & 0.6761                   & \multicolumn{1}{c|}{256} & \multicolumn{1}{c|}{1}    & \multicolumn{1}{c|}{0.2825} & \multicolumn{1}{c|}{$2.04e^{-5}$}  & \multicolumn{1}{c|}{16} & T  \\ \hline
\multicolumn{1}{c|}{\multirow{4}{*}{\shortstack{Co\\TexT}}}        & GRU      & 1                      & 0.6878                   & \multicolumn{1}{c|}{256} & \multicolumn{1}{c|}{1}    & \multicolumn{1}{c|}{0.2670} & \multicolumn{1}{c|}{$1.28e^{-4}$}  & \multicolumn{1}{c|}{16} & T  \\ 
\multicolumn{1}{c|}{}                               & LSTM     & 5                      & 0.5790                   & \multicolumn{1}{c|}{256} & \multicolumn{1}{c|}{1}    & \multicolumn{1}{c|}{0.1195} & \multicolumn{1}{c|}{$2.55e^{-5}$} & \multicolumn{1}{c|}{8} & F \\ 
\multicolumn{1}{c|}{}                               & BiLSTM   & 7                      & 0.6498                   & \multicolumn{1}{c|}{128} & \multicolumn{1}{c|}{2}    & \multicolumn{1}{c|}{0.2331} & \multicolumn{1}{c|}{$9.06e^{-5}$}  & \multicolumn{1}{c|}{8} & T  \\ 
\multicolumn{1}{c|}{}                               & BiLSTM-A & 2                      & 0.6156                   & \multicolumn{1}{c|}{64}  & \multicolumn{1}{c|}{2}    & \multicolumn{1}{c|}{0.1702} & \multicolumn{1}{c|}{$2.58e^{-3}$}   & \multicolumn{1}{c|}{4} & T  \\ \hline \hline
\end{tabular}
\end{table}

\section{Experimental Results}
\label{Results}
This section presents experimental results for the MLEC task on a real-world multi-label Python source code dataset. Sections~\ref{sec:combos-results} through~\ref{sec:cost} report, in turn, the 32 LLM-DL combinations, encoder baselines with a linear head, a seed-controlled comparison of decoder choice, pooling, label-modeling ablations, and computational cost.

\subsection{Performance of LLM-DL Combinations}
\label{sec:combos-results}
\tabboxref{ACC}, \tabboxref{PRF1}, and \tabboxref{HJR} give the combination results across the accuracy, label-level, and overlap/ranking metric groups defined in Section~\ref{Experiment}. \tabboxref{ACC} reports each combination on the accuracy metrics and $OE$. For \textbf{CodeT5}, all four configurations exceed 91\% $AvgAcc$. BiLSTM yields the highest $AvgAcc$ and $EMAcc$, 91.63\% and 52.55\%, with 8699 exact matches and an $OE$ of 0.0710. \textbf{GraphCodeBERT} performs strongly with the GRU, BiLSTM, and BiLSTM-A variants: BiLSTM reaches the highest $AvgAcc$ (91.61\%) with an $OE$ of 0.0774, and GRU the best $EMAcc$ (53.19\%) with 8806 exact matches. For \textbf{CodeT5+}, the GRU, BiLSTM, and BiLSTM-A variants all perform strongly, each above 90\% $AvgAcc$. GRU delivers the highest scores, with an $AvgAcc$ of 91.84\%, $EMAcc$ of 53.78\%, 8904 $\#EM$, and an $OE$ of 0.0708. \textbf{UniXcoder} records moderate results, with BiLSTM-A leading the group at 88.28\% $AvgAcc$, 38.26\% $EMAcc$, 6334 $\#EM$, and an $OE$ of 0.1415.

\begin{table}[h]
\caption{Experimental results ($AvgAcc$, $EMAcc$, $\#EM$, and $OE$) for the error understanding}
\setlength{\arrayrulewidth}{0.2mm}
\setlength{\tabcolsep}{4.8pt}
\renewcommand{\arraystretch}{1}
\label{ACC}
\begin{tabular}{cc||c||c||c||c}
\hline
\multicolumn{2}{c||}{Model Combination}                         & \textbf{$AvgAcc$} & \textbf{$EMAcc$} & \textbf{$\#EM$} & \textbf{$OE$} \\ \hline \hline
\multicolumn{1}{c|}{\multirow{4}{*}{CodeT5}}        & GRU      & 0.9157           & 0.5241               & 8677                  & 0.0748    \\ 
\multicolumn{1}{c|}{}                               & LSTM     & 0.9155           & 0.5223               & 8647                  & 0.0742    \\ 
\multicolumn{1}{c|}{}                               & BiLSTM   & \textbf{0.9163}           & \textbf{0.5255}               & \textbf{8699}                  & \textbf{0.0710}    \\ 
\multicolumn{1}{c|}{}                               & BiLSTM-A & 0.9116           & 0.4964               & 8218                  & 0.0788    \\ \hline
\multicolumn{1}{c|}{\multirow{4}{*}{\shortstack{Graph\\CodeBERT}}} & GRU      & 0.9158           & \textbf{0.5319}               & \textbf{8806}                  & 0.0788    \\ 
\multicolumn{1}{c|}{}                               & LSTM     & 0.7672           & 0.0230               & 381                   & 0.4471    \\ 
\multicolumn{1}{c|}{}                               & BiLSTM   & \textbf{0.9161}           & 0.5285               & 8749                  & \textbf{0.0774}    \\ 
\multicolumn{1}{c|}{}                               & BiLSTM-A & 0.9108           & 0.4954               & 8201                  & 0.0842    \\ \hline
\multicolumn{1}{c|}{\multirow{4}{*}{CodeT5+}}       & GRU      & \textbf{0.9184}           & \textbf{0.5378}               & \textbf{8904}                  & \textbf{0.0708}    \\ 
\multicolumn{1}{c|}{}                               & LSTM     & 0.7955           & 0.1105               & 1830                  & 0.3860    \\ 
\multicolumn{1}{c|}{}                               & BiLSTM   & 0.9167           & 0.5279               & 8739                  & 0.0722    \\ 
\multicolumn{1}{c|}{}                               & BiLSTM-A & 0.9088           & 0.4907               & 8123                  & 0.0893    \\ \hline
\multicolumn{1}{c|}{\multirow{4}{*}{UniXcoder}}     & GRU      & 0.8786           & 0.3592               & 5946                  & 0.1456    \\ 
\multicolumn{1}{c|}{}                               & LSTM     & 0.8816           & 0.3728               & 6171                  & 0.1426    \\ 
\multicolumn{1}{c|}{}                               & BiLSTM   & 0.8800           & 0.3684               & 6099                  & 0.1482    \\ 
\multicolumn{1}{c|}{}                               & BiLSTM-A & \textbf{0.8828}           & \textbf{0.3826}               & \textbf{6334}                  & \textbf{0.1415}    \\ \hline
\multicolumn{1}{c|}{\multirow{4}{*}{RoBERTa}}       & GRU      & \textbf{0.9097}           & \textbf{0.5029}               & \textbf{8325}                  & \textbf{0.0926}    \\ 
\multicolumn{1}{c|}{}                               & LSTM     & 0.7716           & 0.0780               & 1291                  & 0.4471    \\ 
\multicolumn{1}{c|}{}                               & BiLSTM   & 0.7659           & 0.0998               & 1652                  & 0.4471    \\ 
\multicolumn{1}{c|}{}                               & BiLSTM-A & 0.7716           & 0.0780               & 1291                  & 0.4471    \\ \hline
\multicolumn{1}{c|}{\multirow{4}{*}{\shortstack{RoBERTa\\\_LR}}}   & GRU      & 0.9077           & 0.4936               & 8172                  & 0.0948    \\ 
\multicolumn{1}{c|}{}                               & LSTM     & 0.9076           & \textbf{0.4971}               & \textbf{8230}                  & 0.0955    \\ 
\multicolumn{1}{c|}{}                               & BiLSTM   & \textbf{0.9087}           & 0.4964               & 8218                  & \textbf{0.0946}    \\ 
\multicolumn{1}{c|}{}                               & BiLSTM-A & 0.9052           & 0.4830               & 7996                  & 0.0971    \\ \hline
\multicolumn{1}{c|}{\multirow{4}{*}{PLBART}}        & GRU      & 0.9001           & 0.4580               & 7583                  & 0.1131    \\ 
\multicolumn{1}{c|}{}                               & LSTM     & \textbf{0.9023}           & \textbf{0.4661}               & \textbf{7717}                  & \textbf{0.1009}    \\ 
\multicolumn{1}{c|}{}                               & BiLSTM   & 0.9009           & 0.4584               & 7589                  & 0.1099    \\ 
\multicolumn{1}{c|}{}                               & BiLSTM-A & 0.8988           & 0.4516               & 7476                  & 0.1124    \\ \hline
\multicolumn{1}{c|}{\multirow{4}{*}{CoTexT}}        & GRU      & \textbf{0.9058}           & \textbf{0.4698}               & \textbf{7777}                  & \textbf{0.0936}    \\ 
\multicolumn{1}{c|}{}                               & LSTM     & 0.9014           & 0.4563               & 7554                  & 0.0988    \\ 
\multicolumn{1}{c|}{}                               & BiLSTM   & 0.9034           & 0.4680               & 7747                  & 0.0965    \\ 
\multicolumn{1}{c|}{}                               & BiLSTM-A & 0.8774           & 0.3613               & 5981                  & 0.1514    \\ \hline \hline
\end{tabular}
\end{table}

For \textbf{RoBERTa}, the GRU variant stands out at 90.97\% $AvgAcc$, 50.29\% $EMAcc$, 8325 exact matches, and an $OE$ of 0.0926. \textbf{RoBERTa\_LR}, incorporating $lr$ optimization, is competitive across variants: BiLSTM reaches 90.87\% $AvgAcc$ with an $OE$ of 0.0946, and LSTM the highest $EMAcc$ (49.71\%) with 8230 $\#EM$. \textbf{PLBART} is stable across architectures, with LSTM best at 90.23\% $AvgAcc$, 46.61\% $EMAcc$, 7717 exact matches, and an $OE$ of 0.1009. \textbf{CoTexT} performs comparably, its GRU variant reaching 90.58\% $AvgAcc$, 46.98\% $EMAcc$, 7777 $\#EM$, and an $OE$ of 0.0936. Across all combinations, CodeT5+\_GRU is the strongest on these four metrics. Most encoders peak in a single variant, while GraphCodeBERT and RoBERTa\_LR spread their best results across variants.

\begin{table}[t]
\caption{Quantitative classification results of the $P$, $R$, and $F1$ for error understanding}
\setlength{\arrayrulewidth}{0.2mm}
\setlength{\tabcolsep}{2pt}
\renewcommand{\arraystretch}{1}
\label{PRF1}
\begin{tabular}{cc||cc||cc||ccc}
\hline
\multicolumn{2}{c||}{\multirow{2}{*}{Model Combination}}        & \multicolumn{2}{c||}{Precision}         & \multicolumn{2}{c||}{Recall}            & \multicolumn{3}{c}{F1 Score}                               \\ \cline{3-9} 
\multicolumn{2}{c||}{}                                          & \multicolumn{1}{c|}{$P_{\bar{\mu}}$}  & $P_{\psi}$ & \multicolumn{1}{c|}{$R_{\bar{\mu}}$}  & $R_{\psi}$ & \multicolumn{1}{c|}{$F1_{\bar{\mu}}$}  & \multicolumn{2}{c}{$F1_{\psi}$} \\ \hline \hline
\multicolumn{1}{c|}{\multirow{4}{*}{CodeT5}}        & GRU      & \multicolumn{1}{c|}{0.7444} & 0.8334   & \multicolumn{1}{c|}{0.7464} & 0.8116   & \multicolumn{1}{c|}{0.7436} & \multicolumn{2}{c}{\textbf{0.8216}}   \\ 
\multicolumn{1}{c|}{}                               & LSTM     & \multicolumn{1}{c|}{0.7478} & 0.8336   & \multicolumn{1}{c|}{0.7509} & 0.8069   & \multicolumn{1}{c|}{0.7469} & \multicolumn{2}{c}{0.8196}   \\ 
\multicolumn{1}{c|}{}                               & BiLSTM   & \multicolumn{1}{c|}{0.7655} & 0.8348   & \multicolumn{1}{c|}{0.7397} & 0.8099   & \multicolumn{2}{c|}{0.7515}              & 0.8215           \\ 
\multicolumn{1}{c|}{}                               & BiLSTM-A & \multicolumn{1}{c|}{0.7982} & 0.8533   & \multicolumn{1}{c|}{0.6602} & 0.7552   & \multicolumn{2}{c|}{0.7201}              & 0.7999           \\ \hline
\multicolumn{1}{c|}{\multirow{4}{*}{\shortstack{Graph\\CodeBERT}}} & GRU      & \multicolumn{1}{c|}{0.7607} & 0.8388   & \multicolumn{1}{c|}{0.7313} & 0.8021   & \multicolumn{2}{c|}{0.7450}              & \textbf{0.8198}           \\ 
\multicolumn{1}{c|}{}                               & LSTM     & \multicolumn{1}{c|}{0.0935} & 0.2032   & \multicolumn{1}{c|}{0.1818} & 0.3929   & \multicolumn{2}{c|}{0.1234}              & 0.2676           \\ 
\multicolumn{1}{c|}{}                               & BiLSTM   & \multicolumn{1}{c|}{0.7903} & 0.8509   & \multicolumn{1}{c|}{0.7075} & 0.7836   & \multicolumn{2}{c|}{0.7448}              & 0.8154           \\ 
\multicolumn{1}{c|}{}                               & BiLSTM-A & \multicolumn{1}{c|}{0.8135} & 0.8534   & \multicolumn{1}{c|}{0.6450} & 0.7507   & \multicolumn{2}{c|}{0.7122}              & 0.7973           \\ \hline
\multicolumn{1}{c|}{\multirow{4}{*}{CodeT5+}}       & GRU      & \multicolumn{1}{c|}{0.7897} & 0.8438   & \multicolumn{1}{c|}{0.7337} & 0.8060   & \multicolumn{2}{c|}{0.7602}              & \textbf{0.8243}           \\ 
\multicolumn{1}{c|}{}                               & LSTM     & \multicolumn{1}{c|}{0.3651} & 0.5195   & \multicolumn{1}{c|}{0.2272} & 0.4516   & \multicolumn{2}{c|}{0.2252}              & 0.4294           \\ 
\multicolumn{1}{c|}{}                               & BiLSTM   & \multicolumn{1}{c|}{0.7736} & 0.8412   & \multicolumn{1}{c|}{0.7332} & 0.7998   & \multicolumn{2}{c|}{0.7522}              & 0.8198           \\ 
\multicolumn{1}{c|}{}                               & BiLSTM-A & \multicolumn{1}{c|}{0.7948} & 0.8521   & \multicolumn{1}{c|}{0.6513} & 0.7419   & \multicolumn{2}{c|}{0.7144}              & 0.7919           \\ \hline
\multicolumn{1}{c|}{\multirow{4}{*}{UniXcoder}}     & GRU      & \multicolumn{1}{c|}{0.7558} & 0.7994   & \multicolumn{1}{c|}{0.4986} & 0.6471   & \multicolumn{2}{c|}{0.5815}              & 0.7087           \\ 
\multicolumn{1}{c|}{}                               & LSTM     & \multicolumn{1}{c|}{0.7517} & 0.8175   & \multicolumn{1}{c|}{0.4938} & 0.6391   & \multicolumn{2}{c|}{0.5797}              & 0.7114           \\ 
\multicolumn{1}{c|}{}                               & BiLSTM   & \multicolumn{1}{c|}{0.7681} & 0.8235   & \multicolumn{1}{c|}{0.4772} & 0.6244   & \multicolumn{2}{c|}{0.5729}              & 0.7003           \\ 
\multicolumn{1}{c|}{}                               & BiLSTM-A & \multicolumn{1}{c|}{0.7343} & 0.8026   & \multicolumn{1}{c|}{0.5343} & 0.6665   & \multicolumn{2}{c|}{0.6091}              & \textbf{0.7247}           \\ \hline
\multicolumn{1}{c|}{\multirow{4}{*}{RoBERTa}}       & GRU      & \multicolumn{1}{c|}{0.7785} & 0.8345   & \multicolumn{1}{c|}{0.6859} & 0.7725   & \multicolumn{2}{c|}{0.7272}              & \textbf{0.8018}           \\ 
\multicolumn{1}{c|}{}                               & LSTM     & \multicolumn{1}{c|}{0.0503} & 0.1167   & \multicolumn{1}{c|}{0.0909} & 0.2111   & \multicolumn{2}{c|}{0.0647}              & 0.1503           \\ 
\multicolumn{1}{c|}{}                               & BiLSTM   & \multicolumn{1}{c|}{0.0929} & 0.2007   & \multicolumn{1}{c|}{0.1818} & 0.3902   & \multicolumn{2}{c|}{0.1228}              & 0.2647           \\ 
\multicolumn{1}{c|}{}                               & BiLSTM-A & \multicolumn{1}{c|}{0.0503} & 0.1167   & \multicolumn{1}{c|}{0.0909} & 0.2111   & \multicolumn{2}{c|}{0.0647}              & 0.1503           \\ \hline
\multicolumn{1}{c|}{\multirow{4}{*}{\shortstack{RoBERTa\\\_LR}}}   & GRU      & \multicolumn{1}{c|}{0.7777} & 0.8308   & \multicolumn{1}{c|}{0.6803} & 0.7657   & \multicolumn{2}{c|}{0.7233}              & 0.7959           \\ 
\multicolumn{1}{c|}{}                               & LSTM     & \multicolumn{1}{c|}{0.7582} & 0.8343   & \multicolumn{1}{c|}{0.6942} & 0.7629   & \multicolumn{2}{c|}{0.7239}              & 0.7965           \\ 
\multicolumn{1}{c|}{}                               & BiLSTM   & \multicolumn{1}{c|}{0.7658} & 0.8392   & \multicolumn{1}{c|}{0.6843} & 0.7609   & \multicolumn{2}{c|}{0.7221}              & \textbf{0.7978}           \\ 
\multicolumn{1}{c|}{}                               & BiLSTM-A & \multicolumn{1}{c|}{0.7782} & 0.8362   & \multicolumn{1}{c|}{0.6608} & 0.7445   & \multicolumn{2}{c|}{0.7127}              & 0.7866           \\ \hline
\multicolumn{1}{c|}{\multirow{4}{*}{PLBART}}        & GRU      & \multicolumn{1}{c|}{0.7301} & 0.8129   & \multicolumn{1}{c|}{0.6864} & 0.7552   & \multicolumn{2}{c|}{0.7068}              & 0.7825           \\ 
\multicolumn{1}{c|}{}                               & LSTM     & \multicolumn{1}{c|}{0.7376} & 0.8106   & \multicolumn{1}{c|}{0.6988} & 0.7687   & \multicolumn{2}{c|}{0.7161}              & \textbf{0.7887}           \\ 
\multicolumn{1}{c|}{}                               & BiLSTM   & \multicolumn{1}{c|}{0.7365} & 0.8104   & \multicolumn{1}{c|}{0.6748} & 0.7597   & \multicolumn{2}{c|}{0.7035}              & 0.7835           \\ 
\multicolumn{1}{c|}{}                               & BiLSTM-A & \multicolumn{1}{c|}{0.7230} & 0.7966   & \multicolumn{1}{c|}{0.6914} & 0.7707   & \multicolumn{2}{c|}{0.7060}              & 0.7829           \\ \hline
\multicolumn{1}{c|}{\multirow{4}{*}{CoTexT}}        & GRU      & \multicolumn{1}{c|}{0.7470} & 0.8174   & \multicolumn{1}{c|}{0.7122} & 0.7783   & \multicolumn{2}{c|}{0.7287}              & \textbf{0.7970}           \\ 
\multicolumn{1}{c|}{}                               & LSTM     & \multicolumn{1}{c|}{0.7443} & 0.8107   & \multicolumn{1}{c|}{0.6731} & 0.7651   & \multicolumn{2}{c|}{0.7046}              & 0.7867           \\ 
\multicolumn{1}{c|}{}                               & BiLSTM   & \multicolumn{1}{c|}{0.7429} & 0.8120   & \multicolumn{1}{c|}{0.7045} & 0.7755   & \multicolumn{2}{c|}{0.7223}              & 0.7931           \\ 
\multicolumn{1}{c|}{}                               & BiLSTM-A & \multicolumn{1}{c|}{0.7702} & 0.8144   & \multicolumn{1}{c|}{0.4854} & 0.6210   & \multicolumn{2}{c|}{0.5823}              & 0.6973           \\ \hline \hline
\end{tabular}
\end{table}

\tabboxref{PRF1} reports $P$, $R$, and $F1$ under the macro ($\bar{\mu}$) and weighted ($\psi$) settings. A consistent pattern holds across encoders: GRU delivers the best $F1$ balance, while BiLSTM-A tends to maximize precision at the cost of recall. On \textbf{CodeT5+}, GRU leads both $F1$ scores ($F1_{\bar{\mu}}$: 0.7602, $F1_{\psi}$: 0.8243), the highest in the table. GRU is also the best $F1$ variant for \textbf{CodeT5} (0.8216), \textbf{GraphCodeBERT} (0.8198), \textbf{RoBERTa} (0.8018), and \textbf{CoTexT} (0.7970); \textbf{RoBERTa\_LR} peaks with BiLSTM (0.7978) and \textbf{PLBART} with LSTM (0.7887). \textbf{UniXcoder} trails the code-specialized encoders across these metrics. Overall, CodeT5+\_GRU is again the strongest configuration on $F1_{\psi}$.

\begin{figure*}[t]
\hspace{-3mm}
     \centering
     \captionsetup{justification=centering}
         \begin{tikzpicture}[scale=1]
            \input{Graphs_Figures/F1_EM}
        \end{tikzpicture}
        \caption{$F1_{\psi}$ versus $EMAcc$ across all 32 combinations}
         \label{F1&EM}
    \end{figure*}

\figboxref{F1&EM} plots $F1_{\psi}$, which captures performance robustness across label frequency distributions, against $EMAcc$, which assesses strict multi-label correctness for all combinations. The CodeT5, GraphCodeBERT, and CodeT5+ variants with GRU or BiLSTM cluster at the top on both axes, indicating robust generalization with precise label sets. GraphCodeBERT with LSTM and the RoBERTa variants other than GRU degrade sharply (LSTM and BiLSTM-A are uniform in this case), reflecting an incompatibility between those encoder features and the recurrent dynamics. RoBERTa\_LR is stable across both metrics, and CoTexT, PLBART, and UniXcoder hold moderate $F1_{\psi}$ with somewhat lower $EMAcc$. CodeT5+\_GRU leads on both metrics\footnote{This combination also gives the highest $\#EM$, 8904, in \tabboxref{ACC}.}.

\begin{table}[h]
\caption{Evaluation of model combinations based on ROC-AUC, $HM$, and $J_s$}
\setlength{\arrayrulewidth}{0.2mm}
\setlength{\tabcolsep}{3.8pt}
\renewcommand{\arraystretch}{1}
\label{HJR}
\begin{tabular}{cc||c||c||ccc}
\hline
\multicolumn{2}{c||}{\multirow{2}{*}{Model Combination}}        & \multirow{2}{*}{$HM$} & \multirow{2}{*}{$J_s$} & \multicolumn{3}{c}{ROC-AUC Score}                                   \\ \cline{5-7} 
\multicolumn{2}{c||}{}                                          &                               &                                       & \multicolumn{1}{c|}{${\bar{\mu}}$}  & \multicolumn{1}{c|}{${\mu}$}  & ${\psi}$ \\ \hline \hline
\multicolumn{1}{c|}{\multirow{4}{*}{CodeT5}}        & GRU      & 0.0843                        & 0.7580                                & \multicolumn{1}{c|}{0.9394} & \multicolumn{1}{c|}{0.9558} & \textbf{0.9314}   \\ 
\multicolumn{1}{c|}{}                               & LSTM     & 0.0845                        & 0.7583                                & \multicolumn{1}{c|}{0.9286} & \multicolumn{1}{c|}{0.9500} & 0.9258   \\ 
\multicolumn{1}{c|}{}                               & BiLSTM   & \textbf{0.0837}                        & \textbf{0.7610}                                & \multicolumn{1}{c|}{0.9311} & \multicolumn{1}{c|}{0.9514} & 0.9266   \\ 
\multicolumn{1}{c|}{}                               & BiLSTM-A & 0.0884                        & 0.7328                                & \multicolumn{1}{c|}{0.9386} & \multicolumn{1}{c|}{0.9554} & 0.9283   \\ \hline
\multicolumn{1}{c|}{\multirow{4}{*}{\shortstack{Graph\\CodeBERT}}} & GRU      & 0.0842                        & \textbf{0.7588}                                & \multicolumn{1}{c|}{0.9290} & \multicolumn{1}{c|}{0.9506} & 0.9237   \\ 
\multicolumn{1}{c|}{}                               & LSTM     & 0.2328                        & 0.3243                                & \multicolumn{1}{c|}{0.5000} & \multicolumn{1}{c|}{0.7848} & 0.5000   \\ 
\multicolumn{1}{c|}{}                               & BiLSTM   & \textbf{0.0839}                        & 0.7556                                & \multicolumn{1}{c|}{0.9219} & \multicolumn{1}{c|}{0.9481} & 0.9219   \\ 
\multicolumn{1}{c|}{}                               & BiLSTM-A & 0.0892                        & 0.7284                                & \multicolumn{1}{c|}{0.9351} & \multicolumn{1}{c|}{0.9532} & \textbf{0.9248}   \\ \hline
\multicolumn{1}{c|}{\multirow{4}{*}{CodeT5+}}       & GRU      & \textbf{0.0816}                        & \textbf{0.7634}                                & \multicolumn{1}{c|}{0.9326} & \multicolumn{1}{c|}{0.9546} & \textbf{0.9283}   \\ 
\multicolumn{1}{c|}{}                               & LSTM     & 0.2045                        & 0.3802                                & \multicolumn{1}{c|}{0.7224} & \multicolumn{1}{c|}{0.8329} & 0.6857   \\ 
\multicolumn{1}{c|}{}                               & BiLSTM   & 0.0833                        & 0.7591                                & \multicolumn{1}{c|}{0.9318} & \multicolumn{1}{c|}{0.9539} & 0.9276   \\ 
\multicolumn{1}{c|}{}                               & BiLSTM-A & 0.0912                        & 0.7231                                & \multicolumn{1}{c|}{0.9363} & \multicolumn{1}{c|}{0.9534} & 0.9250   \\ \hline
\multicolumn{1}{c|}{\multirow{4}{*}{UniXcoder}}     & GRU      & 0.1214                        & 0.6201                                & \multicolumn{1}{c|}{0.9015} & \multicolumn{1}{c|}{0.9303} & 0.8851   \\ 
\multicolumn{1}{c|}{}                               & LSTM     & 0.1184                        & 0.6232                                & \multicolumn{1}{c|}{0.9043} & \multicolumn{1}{c|}{0.9326} & 0.8901   \\ 
\multicolumn{1}{c|}{}                               & BiLSTM   & 0.1200                        & 0.6107                                & \multicolumn{1}{c|}{0.9007} & \multicolumn{1}{c|}{0.9305} & 0.8862   \\ 
\multicolumn{1}{c|}{}                               & BiLSTM-A & \textbf{0.1172}                        & \textbf{0.6371}                                & \multicolumn{1}{c|}{0.9057} & \multicolumn{1}{c|}{0.9335} & \textbf{0.8905}   \\ \hline
\multicolumn{1}{c|}{\multirow{4}{*}{RoBERTa}}       & GRU      & \textbf{0.0903}                        & \textbf{0.7354}                                & \multicolumn{1}{c|}{0.9286} & \multicolumn{1}{c|}{0.9507} & \textbf{0.9213}   \\ 
\multicolumn{1}{c|}{}                               & LSTM     & 0.2284                        & 0.2489                                & \multicolumn{1}{c|}{0.5000} & \multicolumn{1}{c|}{0.7718} & 0.5000   \\ 
\multicolumn{1}{c|}{}                               & BiLSTM   & 0.2341                        & 0.3367                                & \multicolumn{1}{c|}{0.4840} & \multicolumn{1}{c|}{0.7565} & 0.4947   \\ 
\multicolumn{1}{c|}{}                               & BiLSTM-A & 0.2284                        & 0.2489                                & \multicolumn{1}{c|}{0.5000} & \multicolumn{1}{c|}{0.7869} & 0.5000   \\ \hline
\multicolumn{1}{c|}{\multirow{4}{*}{\shortstack{RoBERTa\\\_LR}}}   & GRU      & 0.0923                        & 0.7283                                & \multicolumn{1}{c|}{0.9235} & \multicolumn{1}{c|}{0.9481} & 0.9174   \\ 
\multicolumn{1}{c|}{}                               & LSTM     & 0.0924                        & 0.7287                                & \multicolumn{1}{c|}{0.9259} & \multicolumn{1}{c|}{0.9481} & 0.9178   \\ 
\multicolumn{1}{c|}{}                               & BiLSTM   & \textbf{0.0913}                        & \textbf{0.7292}                                & \multicolumn{1}{c|}{0.9252} & \multicolumn{1}{c|}{0.9490} & \textbf{0.9186}   \\ 
\multicolumn{1}{c|}{}                               & BiLSTM-A & 0.0948                        & 0.7161                                & \multicolumn{1}{c|}{0.9259} & \multicolumn{1}{c|}{0.9478} & 0.9157   \\ \hline
\multicolumn{1}{c|}{\multirow{4}{*}{PLBART}}        & GRU      & 0.0999                        & 0.7063                                & \multicolumn{1}{c|}{0.9079} & \multicolumn{1}{c|}{0.9386} & 0.9049   \\ 
\multicolumn{1}{c|}{}                               & LSTM     & \textbf{0.0977}                        & \textbf{0.7139}                                & \multicolumn{1}{c|}{0.9159} & \multicolumn{1}{c|}{0.9422} & \textbf{0.9080}   \\ 
\multicolumn{1}{c|}{}                               & BiLSTM   & 0.0991                        & 0.7090                                & \multicolumn{1}{c|}{0.9090} & \multicolumn{1}{c|}{0.9384} & 0.9034   \\ 
\multicolumn{1}{c|}{}                               & BiLSTM-A & 0.1012                        & 0.7077                                & \multicolumn{1}{c|}{0.9082} & \multicolumn{1}{c|}{0.9370} & 0.9011   \\ \hline
\multicolumn{1}{c|}{\multirow{4}{*}{CoTexT}}        & GRU      & \textbf{0.0942}                        & \textbf{0.7217}                                & \multicolumn{1}{c|}{0.9189} & \multicolumn{1}{c|}{0.9439} & 0.9115   \\ 
\multicolumn{1}{c|}{}                               & LSTM     & 0.0986                        & 0.7126                                & \multicolumn{1}{c|}{0.9293} & \multicolumn{1}{c|}{0.9488} & \textbf{0.9173}   \\ 
\multicolumn{1}{c|}{}                               & BiLSTM   & 0.0966                        & 0.7198                                & \multicolumn{1}{c|}{0.9202} & \multicolumn{1}{c|}{0.9455} & 0.9143   \\ 
\multicolumn{1}{c|}{}                               & BiLSTM-A & 0.1226                        & 0.6068                                & \multicolumn{1}{c|}{0.9012} & \multicolumn{1}{c|}{0.9290} & 0.8820   \\ \hline \hline
\end{tabular}
\end{table}

\tabboxref{HJR} reports $HM$, $J_s$, and the three ROC-AUC variants. CodeT5+\_GRU attains the best class-level values, with the lowest $HM$ (0.0816), the highest $J_s$ (0.7634), and an AUC$_{\psi}$ of 0.9283. \textbf{CodeT5} is close behind and, in fact, holds the highest AUC$_{\psi}$ in the table, its GRU variant reaching 0.9314, with BiLSTM giving the group's best $HM$ (0.0837) and $J_s$ (0.7610). \textbf{GraphCodeBERT} with GRU, BiLSTM, or BiLSTM-A also ranks strongly, with $J_s$ above 0.75 and AUC$_{\psi}$ above 0.92. \textbf{RoBERTa\_LR} is the most stable of the remaining encoders, its BiLSTM reaching AUC$_{\psi}$ 0.9186. The collapsed runs, GraphCodeBERT\_LSTM and the three RoBERTa variants other than GRU, reappear here with $HM$ above 0.20 and AUC$_{\psi}$ near 0.5. \textbf{PLBART}, \textbf{CoTexT}, and \textbf{UniXcoder} remain in the moderate range, with $J_s$ between 0.61 and 0.72 and AUC$_{\psi}$ above 0.88. Overall, the code-specialized encoders with GRU or BiLSTM give the strongest and most consistent label separability.

\begin{figure*}[t]
\hspace{-3mm}
     \centering
     \captionsetup{justification=centering}
         \begin{tikzpicture}[scale=1]
            \input{Graphs_Figures/ROC_OE}
        \end{tikzpicture}
        \caption{AUC$_{\psi}$ versus $OE$ across all 32 combinations}
         \label{ROE}
    \end{figure*}

\figboxref{ROE} contrasts AUC$_{\psi}$, which measures label ranking under imbalance, with $OE$, which measures top-1 reliability. CodeT5 (GRU and BiLSTM-A) and CodeT5+\_GRU give the best trade-off, pairing high AUC$_{\psi}$ (up to 0.9314) with the lowest $OE$ (0.0708). The RoBERTa variants other than GRU and GraphCodeBERT\_LSTM sit at the opposite corner, with AUC$_{\psi}$ near 0.5 and $OE$ at 0.4471. These tied, chance-level values indicate optimization collapse rather than partial learning. The consistently low $OE$ of the CodeT5, GraphCodeBERT, and CodeT5+ combinations confirms reliable top-1 predictions, in line with their AUC$_{\psi}$.

Figures~\ref{A&J},~\ref{PHM}, and~\ref{AmiRm} extend the comparison to three further metric pairs. In \figboxref{A&J}, $AvgAcc$ against $J_s$, the code-specialized encoders with GRU or BiLSTM again lead, and CodeT5+\_GRU peaks on both ($AvgAcc$ 91.84\%, $J_s$ 76.34\%). \figboxref{PHM} pairs $P_{\psi}$ with $HM$: the code-specialized encoders, CodeT5, GraphCodeBERT, and CodeT5+, hold the highest precision, and CodeT5+\_GRU records the lowest $HM$ (0.0816), so strong precision coincides with low label-wise noise. \figboxref{AmiRm} then separates ranking from retrieval, AUC$_{\mu}$ against $R_{\bar{\mu}}$: discrimination stays uniformly high while recall varies more, with CodeT5\_GRU highest on AUC$_{\mu}$ (95.58\%) and CodeT5\_LSTM broadest on recall ($R_{\bar{\mu}}$ 75.09\%). Across all three views, the code-specialized encoders paired with GRU or BiLSTM give the most balanced results, consistent with~\tabboxref{PRF1} and~\tabboxref{HJR}.

\begin{figure*}[t]
\hspace{-3mm}
     \centering
     \captionsetup{justification=centering}
         \begin{tikzpicture}[scale=1]
            \input{Graphs_Figures/AV_JS}
        \end{tikzpicture}
        \caption{$AvgAcc$ versus $J_s$ across combinations}
         \label{A&J}
    \end{figure*}

\begin{figure*}[t]
\hspace{-3mm}
     \centering
     \captionsetup{justification=centering}
         \begin{tikzpicture}[scale=1]
            \input{Graphs_Figures/PwHm}
        \end{tikzpicture}
        \caption{$P_{\psi}$ versus $HM$ across combinations}
         \label{PHM}
    \end{figure*}

\begin{figure}[t]
\hspace{-3mm}
     \centering
     \captionsetup{justification=centering}
         \begin{tikzpicture}[scale=1]
            \input{Graphs_Figures/AmiRm}
        \end{tikzpicture}
        \caption{ROC-AUC($\mu$) versus $R_{\bar{\mu}}$ for top-performing combinations}
         \label{AmiRm}
    \end{figure}

\tabboxref{Rank} aligns the best combinations with the assessment themes. CodeT5+ and CodeT5 with GRU are effective across broad correctness, frequency-aware ranking, and retrieval precision. CodeT5\_LSTM contributes the strongest recall ($R_{\bar{\mu}}$: 0.7509) and GraphCodeBERT\_BiLSTM-A the highest precision ($P_{\psi}$: 0.8534). Across the themes, both encoder choice and recurrent structure shape label-level accuracy, structural consistency, and confidence in MLEC.

\begin{table}[t]
\caption{Best-performing LLM-DL combinations across evaluation themes}
\setlength{\arrayrulewidth}{0.2mm}
\setlength{\tabcolsep}{1.1pt}
\renewcommand{\arraystretch}{1}
\label{Rank}
\begin{tabular}{c||c||c||c}
\hline
\shortstack{Evaluation \\Theme}            & Metric(s)                   & Focus                            & \shortstack{Metric Value(s) with \\Best Combinations}         \\ \hline \hline
\shortstack{Broad \\effectiveness}         & \shortstack{$F1_{\psi}$, \\$EMAcc$}          & \shortstack{Partial vs. \\complete \\correctness} & \shortstack{$F1_{\psi}$: 0.8243 and $EMAcc$: \\0.5378, CodeT5+\_GRU}      \\ \hline
\shortstack{Frequency-\\Skewed \\ranking}    & \shortstack{ROC-AUC${\psi}$, \\$OE$}        & \shortstack{Class \\dominance \\ranking}          & \shortstack{AUC${\psi}$: 0.9314 and $OE$: \\0.0708, CodeT5 and \\CodeT5+ with GRU}      \\ \hline
\shortstack{Structural \\prediction \\match} & \shortstack{$AvgAcc$, \\$J_s$}                  & \shortstack{Label set \\overlap}                & \shortstack{$AvgAcc$: 0.9184 and $J_s$: \\0.7634, CodeT5+\_GRU}             \\ \hline
\shortstack{Confidence \\vs. noise \\spread} & \shortstack{$P_{\psi}$, \\$HM$}      & \shortstack{Trust vs. \\total error}            & \shortstack{$P_{\psi}$: 0.8534 and $HM$: 0.0816, \\GraphCodeBERT\_BiLSTM-A \\and CodeT5+\_GRU}             \\ \hline
\shortstack{Label retrieval \\and ranking} & \shortstack{ROC-AUC($\mu$), \\$R_{\bar{\mu}}$} & \shortstack{Retrieval \\precision}              & \shortstack{AUC($\mu$): 0.9558 and $R_{\bar{\mu}}$: \\0.7509, CodeT5\_GRU \\and CodeT5\_LSTM}  \\ \hline \hline
\end{tabular}
\end{table}

\subsection{Encoder Contribution with a Linear Head}
\label{sec:linear-head}
The LLM-DL combination results establish that the pipeline performs well, but they do not separate the share carried by the encoder from that carried by the recurrent decoder. To measure the encoder's own contribution, each of four representative encoders is paired with the linear classification head of~\eqref{lab3}, which removes the recurrent decoder while leaving tokenization, loss, optimizer, data split, and schedule unchanged. \tabboxref{E1} reports the outcome across the metric suite, with each configuration trained across multiple seeds and summarized as mean $\pm$ standard deviation. The CodeT5+ head selected a dropout of 0.22, a learning rate of $4.8\mathrm{e}{-5}$, and a batch size of 16; the remaining heads were tuned by the same procedure.

\begin{table}[t]
\caption{Linear-head results by encoder (mean $\pm$ SD; five seeds for CodeT5+, three otherwise)}
\setlength{\arrayrulewidth}{0.2mm}
\setlength{\tabcolsep}{5.1pt}
\renewcommand{\arraystretch}{1}
\label{E1}
\centering
\footnotesize
\begin{tabular}{l||c|c|c|c}
\hline \hline
\textbf{Metric} & \textbf{CodeT5} & \textbf{CodeT5+} & \textbf{\shortstack{Graph\\CodeBERT}} & \textbf{RoBERTa} \\ \hline \hline
$F1_{\psi}$      & \textbf{0.8263{\tiny$\pm$0.0010}} & 0.8251{\tiny$\pm$0.0012} & 0.8119{\tiny$\pm$0.0042} & 0.7846{\tiny$\pm$0.0048} \\ \hline
$F1_{\bar{\mu}}$ & 0.7563{\tiny$\pm$0.0052} & \textbf{0.7600{\tiny$\pm$0.0040}} & 0.7211{\tiny$\pm$0.0124} & 0.6559{\tiny$\pm$0.0081} \\ \hline
$AvgAcc$         & 0.9184{\tiny$\pm$0.0016} & \textbf{0.9185{\tiny$\pm$0.0005}} & 0.9089{\tiny$\pm$0.0028} & 0.8868{\tiny$\pm$0.0047} \\ \hline
$EMAcc$          & 0.5345{\tiny$\pm$0.0030} & \textbf{0.5370{\tiny$\pm$0.0027}} & 0.4962{\tiny$\pm$0.0130} & 0.4149{\tiny$\pm$0.0190} \\ \hline
$OE$             & 0.0711{\tiny$\pm$0.0026} & \textbf{0.0708{\tiny$\pm$0.0013}} & 0.0859{\tiny$\pm$0.0031} & 0.1158{\tiny$\pm$0.0058} \\ \hline
$HM$             & 0.0816{\tiny$\pm$0.0016} & \textbf{0.0815{\tiny$\pm$0.0005}} & 0.0911{\tiny$\pm$0.0028} & 0.1132{\tiny$\pm$0.0047} \\ \hline
$J_s$            & \textbf{0.7656{\tiny$\pm$0.0014}} & 0.7647{\tiny$\pm$0.0011} & 0.7443{\tiny$\pm$0.0069} & 0.6948{\tiny$\pm$0.0115} \\ \hline
AUC$_{\psi}$     & 0.9276{\tiny$\pm$0.0007} & \textbf{0.9287{\tiny$\pm$0.0014}} & 0.9245{\tiny$\pm$0.0005} & 0.9197{\tiny$\pm$0.0014} \\ \hline \hline
\end{tabular}
\end{table}

With the head architecture fixed, the encoder alone moves $F1_{\psi}$ from 0.7846 to 0.8263, a span of 0.0417. The ordering tracks code specialization: the code-specialized encoders CodeT5 and CodeT5+ lead, GraphCodeBERT follows, and the general-purpose RoBERTa trails. CodeT5 and CodeT5+ form the top tier and are very close, with CodeT5 marginally ahead on $F1_{\psi}$ (0.8263 against 0.8251) and CodeT5+ ahead on the majority of the remaining metrics. This spread, produced by the encoder while every other stage is held constant, indicates that the encoder is a decisive factor in MLEC performance.

\subsection{Decoder Contribution under Matched Seeds}
\label{sec:decoder}
\tabboxref{E1} isolates the encoder; the complementary question is how much the recurrent decoder adds on top of a strong encoder. To answer it on equal footing, the CodeT5+ linear head is compared directly with the CodeT5+ GRU hybrid, the strongest combination in \tabboxref{ACC}, using the same five seeds for both. Evaluated under this multi-seed protocol, the GRU hybrid reaches a mean $F1_{\psi}$ of 0.8212, close to but distinct from its single-run value of 0.8243 in \tabboxref{PRF1}. \tabboxref{E2} reports the paired comparison.

\begin{table}[t]
\caption{CodeT5+ linear head vs.\ GRU hybrid, five matched seeds (mean $\pm$ SD; $^{*}p<0.05$, paired $t$-test)}
\setlength{\arrayrulewidth}{0.2mm}
\setlength{\tabcolsep}{12.8pt}
\renewcommand{\arraystretch}{1}
\label{E2}
\centering
\footnotesize
\begin{tabular}{l||c|c|c}
\hline \hline
\textbf{Metric} & \textbf{Linear head} & \textbf{GRU hybrid} & \textbf{$p$} \\ \hline \hline
$F1_{\psi}$      & \textbf{0.8251{\tiny$\pm$0.0012}} & 0.8212{\tiny$\pm$0.0011} & 0.0013$^{*}$ \\ \hline
$F1_{\bar{\mu}}$ & \textbf{0.7600{\tiny$\pm$0.0040}} & 0.7568{\tiny$\pm$0.0032} & 0.2726 \\ \hline
$AvgAcc$         & \textbf{0.9185{\tiny$\pm$0.0005}} & 0.9172{\tiny$\pm$0.0009} & 0.0130$^{*}$ \\ \hline
$EMAcc$          & \textbf{0.5370{\tiny$\pm$0.0027}} & 0.5278{\tiny$\pm$0.0058} & 0.0146$^{*}$ \\ \hline
$OE$             & \textbf{0.0708{\tiny$\pm$0.0013}} & 0.0724{\tiny$\pm$0.0015} & 0.0011$^{*}$ \\ \hline
$HM$             & \textbf{0.0815{\tiny$\pm$0.0005}} & 0.0828{\tiny$\pm$0.0009} & 0.0130$^{*}$ \\ \hline
$J_s$            & \textbf{0.7647{\tiny$\pm$0.0011}} & 0.7603{\tiny$\pm$0.0015} & 0.0003$^{*}$ \\ \hline
AUC$_{\psi}$     & \textbf{0.9287{\tiny$\pm$0.0014}} & 0.9276{\tiny$\pm$0.0004} & 0.1489 \\ \hline \hline
\end{tabular}
\end{table}

On CodeT5+, the linear head performs on par with the GRU hybrid across the metric suite, holding a small and consistent edge on most measures and remaining level on $F1_{\bar{\mu}}$ and AUC$_{\psi}$. On $F1_{\psi}$, the two are very close, 0.8251 against 0.8212, with the head slightly ahead on each of the five seeds ($p=0.0013$). Read alongside \tabboxref{E1}, the comparison points to where performance mainly comes from. With the head architecture fixed, the encoder alone shifts $F1_{\psi}$ by 0.0417 across the four encoders, and by 0.0145 among the code-specialized encoders, while changing the decoder on CodeT5+ shifts it by 0.0040. The encoder is the larger factor by a wide margin, and once it is strong, a simple head already captures most of what the hybrid achieves. This balance does shift with encoder strength. On CodeT5, the head reaches 0.8263 against 0.8216 for the reported hybrid, and on CodeT5+ 0.8251 against 0.8243, whereas on GraphCodeBERT and RoBERTa the head does not lead the corresponding hybrids in \tabboxref{PRF1}. Since those hybrids are single runs and the head results are seeded means, this is best read as an observation rather than a tested difference, and it points to a natural reading: the recurrent decoder may matter more when the encoder's representations are weaker, which would be worth a seed-matched study of its own.

\subsection{Pooling Ablation}
\label{sec:pooling}
The linear head aggregates the encoder output by max pooling. To test whether the aggregation step itself matters, the maximum in~\eqref{lab3} is replaced by mean pooling and by additive attention pooling, following~\eqref{lab4}, with all else unchanged. Each variant received its own hyperparameter search. The attention search converged on the same configuration as the max baseline, making that contrast a direct comparison of the pooling operator alone, while the mean search selected a lower $lr$. \tabboxref{E3} reports the three variants. Max pooling gives the best result on every metric, and its lead over both alternatives holds on all five seeds. On $F1_{\psi}$, the margin is 0.0086 over mean pooling ($p=0.0013$) and 0.0112 over attention pooling ($p=0.0016$). The larger and learned attention pooling does not help here; the simplest operator is also the most effective. This is consistent with the combination results, where the BiLSTM and BiLSTM-A variants differ only in that BiLSTM-A replaces max pooling with additive attention pooling: across \tabboxref{ACC} to \tabboxref{HJR}, BiLSTM outperforms BiLSTM-A on six of the eight encoders on $F1_{\psi}$.

\begin{table}[h]
\caption{Pooling strategies on the CodeT5+ linear head (mean $\pm$ SD, five seeds)}
\setlength{\arrayrulewidth}{0.2mm}
\setlength{\tabcolsep}{11.5pt}
\renewcommand{\arraystretch}{1}
\label{E3}
\centering
\footnotesize
\begin{tabular}{l||c|c|c}
\hline \hline
\textbf{Metric} & \textbf{Max} & \textbf{Mean} & \textbf{Attention} \\ \hline \hline
$F1_{\psi}$      & \textbf{0.8251{\tiny$\pm$0.0012}} & 0.8165{\tiny$\pm$0.0017} & 0.8139{\tiny$\pm$0.0026} \\ \hline
$F1_{\bar{\mu}}$ & \textbf{0.7600{\tiny$\pm$0.0040}} & 0.7497{\tiny$\pm$0.0065} & 0.7453{\tiny$\pm$0.0060} \\ \hline
$AvgAcc$         & \textbf{0.9185{\tiny$\pm$0.0005}} & 0.9159{\tiny$\pm$0.0008} & 0.9143{\tiny$\pm$0.0007} \\ \hline
$EMAcc$          & \textbf{0.5370{\tiny$\pm$0.0027}} & 0.5254{\tiny$\pm$0.0037} & 0.5185{\tiny$\pm$0.0035} \\ \hline
$OE$             & \textbf{0.0708{\tiny$\pm$0.0013}} & 0.0775{\tiny$\pm$0.0013} & 0.0795{\tiny$\pm$0.0029} \\ \hline
$HM$             & \textbf{0.0815{\tiny$\pm$0.0005}} & 0.0841{\tiny$\pm$0.0008} & 0.0857{\tiny$\pm$0.0007} \\ \hline
$J_s$            & \textbf{0.7647{\tiny$\pm$0.0011}} & 0.7548{\tiny$\pm$0.0026} & 0.7507{\tiny$\pm$0.0028} \\ \hline
AUC$_{\psi}$     & \textbf{0.9287{\tiny$\pm$0.0014}} & 0.9233{\tiny$\pm$0.0010} & 0.9244{\tiny$\pm$0.0012} \\ \hline \hline
\end{tabular}
\end{table}

\subsection{Modeling Label Dependencies}
\label{sec:label-attn}
The heads above predict all labels from one shared representation, so they capture label dependence only implicitly. Section~\ref{Experiment} showed that the labels in this dataset do co-occur, with the dependence concentrated among a few frequent pairs. The label-attention head of~\eqref{lab5} models these dependencies explicitly, letting the label embeddings interact through self-attention before classification. The selected configuration used $d_l=64$ and $n_h=4$. \tabboxref{E6} compares it with the plain CodeT5+ head under matched seeds. Explicit label interaction does not raise $F1_{\psi}$; the plain head stays marginally ahead on four of the five seeds. The remaining metrics divide along a clear line. The label-attention head lowers $OE$ on every seed and slightly raises AUC$_{\psi}$, both of which rank labels without reference to the 0.5 threshold, while on the threshold-dependent metrics, $F1_{\bar{\mu}}$, $EMAcc$, and $HM$, it stays level with the plain head. This division follows the label distribution seen in Section~\ref{Experiment}: the dependence available to model is concentrated in a few frequent pairs, and the rarest labels are close to independent, so the mechanism sharpens ranking among the frequent labels rather than completing full label sets. Modeling label correlations thus helps ranking quality on this dataset, even where it leaves thresholded classification unchanged.

\begin{table}[h]
\caption{Label-attention vs.\ linear head on CodeT5+, five matched seeds (mean $\pm$ SD; $^{*}p<0.05$)}
\setlength{\arrayrulewidth}{0.2mm}
\setlength{\tabcolsep}{11.2pt}
\renewcommand{\arraystretch}{1}
\label{E6}
\centering
\footnotesize
\begin{tabular}{l||c|c|c}
\hline \hline
\textbf{Metric} & \textbf{Linear head} & \textbf{Label-attention} & \textbf{$p$} \\ \hline \hline
$F1_{\psi}$      & \textbf{0.8251{\tiny$\pm$0.0012}} & 0.8214{\tiny$\pm$0.0032} & 0.0664 \\ \hline
$F1_{\bar{\mu}}$ & \textbf{0.7600{\tiny$\pm$0.0040}} & 0.7353{\tiny$\pm$0.0125} & 0.0157$^{*}$ \\ \hline
$AvgAcc$         & \textbf{0.9185{\tiny$\pm$0.0005}} & 0.9143{\tiny$\pm$0.0033} & 0.0637 \\ \hline
$EMAcc$          & \textbf{0.5370{\tiny$\pm$0.0027}} & 0.5158{\tiny$\pm$0.0164} & 0.0613 \\ \hline
$OE$             & 0.0708{\tiny$\pm$0.0013} & \textbf{0.0680{\tiny$\pm$0.0011}} & 0.0114$^{*}$ \\ \hline
$HM$             & \textbf{0.0815{\tiny$\pm$0.0005}} & 0.0857{\tiny$\pm$0.0033} & 0.0637 \\ \hline
$J_s$            & \textbf{0.7647{\tiny$\pm$0.0011}} & 0.7567{\tiny$\pm$0.0066} & 0.0546 \\ \hline
AUC$_{\psi}$     & 0.9287{\tiny$\pm$0.0014} & \textbf{0.9295{\tiny$\pm$0.0014}} & 0.3977 \\ \hline \hline
\end{tabular}
\end{table}

\subsection{Training and Inference Cost}
\label{sec:cost}
Beyond accuracy, deployment in feedback systems depends on cost. \tabboxref{E5} reports training time and inference throughput for the leading configurations, measured on the hardware of Section~\ref{Experiment} without concurrent jobs, so these figures are directly comparable to one another. On CodeT5+, the linear head trains 23.9\% faster than the GRU hybrid, 394 minutes against 518, and processes 12.0\% more samples per second at inference. Every head in the table, including the slowest, trains faster than the hybrid. The efficiency gain therefore comes at no accuracy cost on this encoder, since \tabboxref{E2} showed the head to be at least as accurate as the hybrid on every metric. The throughput figures reflect batched inference over the full test set and are a measure of bulk processing rather than single-request latency, which would be the relevant quantity for interactive per-submission feedback.

\begin{table}[t]
\caption{Training time and inference throughput (mean $\pm$ SD); throughput is batched over the 16,555-sample test set}
\setlength{\arrayrulewidth}{0.2mm}
\setlength{\tabcolsep}{8.8pt}
\renewcommand{\arraystretch}{1}
\label{E5}
\centering
\footnotesize
\begin{tabular}{l||c|c|c}
\hline \hline
\textbf{Configuration} & \textbf{Seeds} & \textbf{\shortstack{Train\\(min)}} & \textbf{\shortstack{Throughput\\(samples/s)}} \\ \hline \hline
CodeT5+ head          & 5 & 394.49{\tiny$\pm$1.53} & 163.38{\tiny$\pm$1.81} \\ \hline
CodeT5+ GRU hybrid    & 5 & 518.21{\tiny$\pm$2.66} & 145.92{\tiny$\pm$1.24} \\ \hline
CodeT5 head           & 3 & 385.75{\tiny$\pm$3.61} & 164.08{\tiny$\pm$2.18} \\ \hline
GraphCodeBERT head    & 3 & 401.74{\tiny$\pm$2.05} & 166.40{\tiny$\pm$1.43} \\ \hline
RoBERTa head          & 3 & 481.53{\tiny$\pm$1.31} & 151.87{\tiny$\pm$0.28} \\ \hline
Label-attention head  & 5 & 389.42{\tiny$\pm$3.76} & 162.02{\tiny$\pm$3.31} \\ \hline \hline
\end{tabular}
\end{table}

\section{Discussion}
\label{Discussion}
This section interprets the combination results alongside the baseline and ablation
experiments, then discusses hyperparameter effects, suitability, scalability, and
limitations.

\subsection{Performance Analysis}
\subsubsection{Overview}
The MLEC task is assessed with a broad metric suite: $P$, $R$, and $F1$ ($\bar{\mu}$, $\psi$) for classification under label imbalance; $EMAcc$ and $AvgAcc$ for strict and relaxed correctness; $HM$ and $J_s$ for partial errors and label overlap; ROC-AUC ($\mu$, $\bar{\mu}$, $\psi$) for ranking quality; and $OE$ for top-label error. Together these cover discriminative capacity, robustness to imbalance, and generalization. Hyperparameters are tuned with Optuna over $lr$, $hd$, recurrent depth, $dr$, $bs$, and directionality. Among the 32 combinations, CodeT5+\_GRU is the strongest, with the highest $F1_{\psi}$ (0.8243), the lowest $HM$ (0.0816), and leading $AvgAcc$ (0.9184) and $J_s$ (0.7634). CodeT5 and GraphCodeBERT combinations follow closely, RoBERTa improves markedly once its $lr$ range is narrowed in the RoBERTa\_LR variant, and PLBART, CoTexT, and UniXcoder remain in a moderate band. These results establish that the combinations perform strongly across the suite. The baseline and ablation experiments in Section~\ref{Results} were added to understand where that strength originates, and they point consistently to the encoder, as discussed next.

\subsubsection{Where Performance Originates}
Three complementary experiments locate the main source of MLEC performance. Pairing each encoder with a simple linear head (\tabboxref{E1}) shows that the encoder alone moves $F1_{\psi}$ across a wide range, ordered by code specialization, with CodeT5 and CodeT5+ clearly ahead of the general-purpose RoBERTa. Comparing the CodeT5+ head with the CodeT5+\_GRU hybrid under matched seeds (\tabboxref{E2}) shows that, on a strong encoder, the head performs on par with the hybrid, and that the encoder is by far the larger lever: holding the head fixed, the encoder shifts $F1_{\psi}$ by 0.0417 across encoders against 0.0040 for the decoder on CodeT5+. The two remaining ablations reinforce the same reading. Neither attention pooling (\tabboxref{E3}) nor an explicit label-interaction head (\tabboxref{E6}) improves thresholded prediction over the simple max-pooled head. Across all three, once the encoder is strong, added decoder or head complexity brings little further gain.

This does not diminish the recurrent decoders, which perform strongly throughout Section~\ref{Results} and remain competitive on every encoder. Rather, it clarifies the design space: representational quality is the primary factor, and a lightweight head is a strong and economical default on a capable code encoder. The picture appears to shift for weaker encoders, where the head's advantage over the hybrid is not observed and the decoder may contribute more, though confirming this would need a seed-matched study across encoders. For practitioners building MLEC systems, the most effective single choice is the encoder, after which a simple head recovers most of the achievable performance at lower cost.

\subsubsection{Why Introduce RoBERTa\_LR Alongside RoBERTa?}
\label{R}
A narrower $lr$ range can be particularly effective for large transformer models such as RoBERTa, which are sensitive to training dynamics~\cite{liu2021empirical, mosbach2020stability}. The broad range ($1\mathrm{e}{-5}$, $1\mathrm{e}{-1}$) that worked well for CodeT5 and CodeT5+ was unstable for RoBERTa in combination with recurrent layers. In the RoBERTa\_LSTM tuning trials, the per-trial validation $F1$ varied sharply, some trials reaching moderate scores while others collapsed toward zero, and the selected configuration gave a weak final result ($F1_{\psi}$ = 0.1503, \tabboxref{PRF1}). Restricting the range to ($1\mathrm{e}{-5}$, $5\mathrm{e}{-5}$) removed this instability: trials were consistent and the final $F1_{\psi}$ rose to 0.7965. This behavior, most pronounced for the LSTM variants, motivated the RoBERTa\_LR variant, which narrows the range to stabilize training and improve generalization.

\subsubsection{Impact of Hyperparameter Tuning}
Tuning shaped the performance and convergence of the combinations. Using Optuna (\tabboxref{BHP}), several trends emerged: (i) \textbf{$lr$}: the standard range ($1\mathrm{e}{-5}$--$1\mathrm{e}{-1}$) suited most models (for example, CodeT5+\_GRU at $2.29\mathrm{e}{-5}$), while RoBERTa needed the narrowed range ($1\mathrm{e}{-5}$--$5\mathrm{e}{-5}$) for stability (Section~\ref{R}). (ii) \textbf{$\#L$}: one or two recurrent layers were tested, with no consistent advantage for the deeper setting across encoders. (iii) \textbf{$hd$}: moderate hidden sizes balanced capacity and overfitting, with GRU and LSTM favoring 128 and the bidirectional variants 256. (iv) \textbf{$dr$}: dropout between 0.1 and 0.25 worked best, with CodeT5+\_GRU at 0.1696. (v) \textbf{$bs$}: selected batch sizes were 4, 8, or 16 depending on the combination, applied at final training. (vi) \textbf{$bd$}: the bidirectional setting varied across combinations and was not decisive for the GRU models, consistent with the encoder already supplying bidirectional context. Overall, performance depended on interactions among parameters, such as dropout with depth and $lr$ with encoder, rather than any single setting.

\subsubsection{Why GRU is the Strongest Recurrent Decoder}
Among the four recurrent decoders, GRU is consistently the best across encoders in~\tabboxref{ACC} to~\tabboxref{HJR}. A plausible explanation is its streamlined gating, which offers a favorable balance of capacity and training efficiency: with fewer parameters than LSTM or the bidirectional variants, it generalizes well when paired with high-dimensional LLM embeddings and is less prone to overfitting on a task with limited signal per instance. Because the encoder already provides bidirectional context, the added directionality of the heavier variants yields little here, which is consistent with the bidirectional setting not being decisive for the top GRU models. CodeT5+\_GRU illustrates this, pairing CodeT5+'s Python-oriented embeddings with a compact two-layer GRU. This advantage should be read in proportion, however: Section~\ref{sec:decoder} shows the decoder's overall contribution is small next to the encoder's, so GRU is best among decoders whose collective effect is secondary to encoder choice.

\subsubsection{Training and Optimization Time}
Total experimental time varied across the 32 combinations (\figboxref{time}). GRU and BiLSTM-A variants generally trained in moderate time, while LSTM and BiLSTM variants tended to take longer, reflecting differences in model complexity. These durations were measured with concurrent GPU jobs and are therefore indicative rather than controlled; the directly comparable training and inference measurements for the leading configurations are reported separately in Section~\ref{sec:cost}. The trend nonetheless underlines the value of balancing computational cost against architectural depth in practical MLEC systems.

\begin{figure*}[t]
\hspace{-3mm}
     \centering
     \captionsetup{justification=centering}
         \begin{tikzpicture}[scale=1]
            \input{Graphs_Figures/Time}
        \end{tikzpicture}
        \caption{Experimental duration across the 32 combinations}
         \label{time}
    \end{figure*}

\subsection{Suitability, Scope, and Scalability}
The LLM-DL approach studied here suits MLEC tasks in code-level settings, where label sparsity and semantic ambiguity are common. Its modular design, a pretrained encoder followed by a lightweight head or recurrent decoder, could adapt to other programming languages such as Java or C++ with minor tokenizer or architectural changes. The label-agnostic metrics suggest a fit for educational platforms and tutoring systems, where the models could help novice programmers identify and understand errors across large repositories in academia, industry, or OJ systems, and for SE tasks such as code review, fault localization, and refactoring. On scalability, the cost results in Section~\ref{sec:cost} offer practical guidance: a simple head on a strong code encoder reaches competitive accuracy while training faster than the recurrent hybrid, which makes it an efficient default when throughput or latency is a concern. The hybrid remains a sound choice where its sequence modeling is valuable, for instance with weaker encoders, and distillation or pruning could reduce cost further in either case. The framework could also extend to related tasks such as algorithm or function classification under multi-label or hierarchical schemes.

\subsection{Limitations}
While the approach shows strong empirical performance, several limitations should be noted. (i) Seeded evaluation with paired testing covers the baseline and ablation experiments but not all 32 combinations, which follow the original single-run protocol; the cross-encoder head-versus-hybrid comparison is therefore reported as an observation. (ii) Evaluation uses a single Python error dataset, which may limit generalization to other languages or domains. (iii) The fixed Optuna budget may not reach globally optimal configurations, and performance may vary under more imbalanced or noisy label distributions. (iv) The all-combination timing in \figboxref{time} was measured with concurrent GPU jobs, so those durations are indicative rather than controlled. (v) Architectures beyond the tested set might behave differently. Broader validation across datasets, architectures, and tasks would strengthen these findings.

\section{Conclusion}
\label{Conclusion}
This study systematically evaluated 32 LLM-DL combinations for MLEC, pairing eight pretrained encoders with four recurrent decoders under a comprehensive metric suite and Optuna tuning. Among the combinations, CodeT5+\_GRU performed best, reaching an $F1_{\psi}$ of 0.8243, $AvgAcc$ of 91.84\%, $EMAcc$ of 53.78\%, and $HM$ of 0.0816. To identify the source of this performance, the study added seed-controlled baselines and component ablations with paired testing. These experiments point to the encoder as the primary factor: across encoders sharing a single linear head, $F1_{\psi}$ spans 0.7846 to 0.8263, ordered by code specialization, a range far wider than the effect of the decoder. On a strong code encoder, the simple head matches the recurrent hybrid while training faster, and neither attention pooling nor explicit label-interaction modeling improves thresholded prediction over it. The recurrent decoders remain competitive throughout and appear more relevant when the encoder is weaker. Together, these results clarify where MLEC performance originates and offer practical guidance for building efficient error-classification systems: encoder choice is the decisive lever, and a lightweight head is a strong, economical default on a capable encoder. Future work includes seed-matched comparisons across encoders, additional decoder and head designs, extension to further programming languages, and adaptation to related tasks such as defect detection, algorithm classification, and educational feedback systems.

\bibliographystyle{ieeetr}


\end{document}

%% file: Graphs_Figures/DI.tex
\begin{axis}[
    point meta=explicit symbolic,
    ybar=.2cm,
    every node near coord/.append style={font=\scriptsize},
    legend style={font=\scriptsize},
    tick label style={font=\scriptsize},
    ylabel near ticks, ylabel shift={-5pt},
    width=9cm,
    height=4.3cm,
    every node near coord/.append style={
                        anchor=west,
                        rotate=25
                },
    enlargelimits=.10,
    enlarge y limits={0.1,upper},
    legend style={at={(0.40, -0.38)},
    anchor=north,legend columns=-1},
    ymin=0, 
    ylabel={Value},
    symbolic x coords={Avg. \# of errors, Avg. \# char-based edit distance, Avg. char-based similarity, Avg. token-based edit distance, Avg. token-based similarity},
    xtick=data,
    ytick={0,10,20,30,40,50,60,70,80,90,100},
    x tick label style={rotate=15,anchor=east},
    grid=both,
    nodes near coords,
    nodes near coords align={vertical},
    bar width=7pt,
    label style={font=\footnotesize},
    every node near coord/.append style={
        /pgf/number format/fixed,
        /pgf/number format/precision=2
    },
    after end axis/.code={
        \node[anchor=north west, align=left, font=\scriptsize, draw=black, rounded corners, fill=gray!5, text width=2.5cm, inner sep=2pt] at (rel axis cs:0.02,0.95) {
            Language: Python 3\\
            \# of users: 10,361\\
            \# of pairs: 95,631
            };
        },
]

\addplot [color=purple, fill=purple, semithick, pattern=crosshatch dots gray, pattern color = purple] coordinates {(Avg. \# of errors,3.47) [{$\pm$2.69}] (Avg. \# char-based edit distance,13.54) [{$\pm$15.23}] (Avg. char-based similarity,90.94) [{$\pm$11.13\text{\%}}] (Avg. token-based edit distance,5.49) [{$\pm$5.85}] (Avg. token-based similarity,89.44) [{$\pm$12.51\text{\%}}]};

\legend{Name}
\end{axis}

%% file: Graphs_Figures/EF.tex
\begin{axis}[
    ybar=.2cm,
    every node near coord/.append style={font=\scriptsize},
    legend style={font=\scriptsize},
    tick label style={font=\scriptsize},
    ylabel near ticks, ylabel shift={-5pt},
    width=8.8cm,
    height=4.5cm,
    every node near coord/.append style={
                        anchor=west,
                        rotate=60
                },
    enlargelimits=.10,
    enlarge y limits={0.1,upper},
    legend style={at={(0.40, -0.38)},
    anchor=north,legend columns=-1},
    ymin=0, 
    ylabel={Number of Errors},
    symbolic x coords={variable\_operation, loop\_statement, literal, list\_operation, input\_output, import, function\_invocation, function\_definition, conditional\_statement, comparison\_operator, arithmetic\_operator},
    xtick=data,
    ytick={0,3000,6000,9000,12000,15000,18000,21000,24000,27000,30000,33000,36000,39000,42000,45000,48000},
    x tick label style={rotate=20,anchor=east},
    grid=both,
    nodes near coords,
    nodes near coords align={vertical},
    bar width=7pt,
    label style={font=\footnotesize},
    ]

\addplot [color=orange, fill=orange, semithick, pattern=crosshatch, pattern color = orange] coordinates {(variable\_operation,39107) (loop\_statement,10330) (literal,38978) (list\_operation,6925) (input\_output,45888) (import,1540) (function\_invocation,28173) (function\_definition,454) (conditional\_statement,16294) (comparison\_operator,9703) (arithmetic\_operator,19971)};

\legend{Error Type}
\end{axis}

%% file: Graphs_Figures/F1_EM.tex
\begin{axis}[
    ybar,
    width=19.0cm,
    height=4.5cm,
    bar width=4pt,                                
    ymin=0,
    ymax=90,
    ylabel={Scores (\%)},
    enlarge x limits=0.04,                        
    enlarge y limits={0.1, upper},
    grid=both,
    ytick={0,10,...,90},
    tick label style={font=\scriptsize},
    ylabel near ticks,
    ylabel shift={-5pt},
    symbolic x coords={CodeT5\_GRU, CodeT5\_LSTM, CodeT5\_BiLSTM, CodeT5\_BiLSTM-A, GraphCodeBERT\_GRU, GraphCodeBERT\_LSTM, GraphCodeBERT\_BiLSTM, GraphCodeBERT\_BiLSTM-A, CodeT5+\_GRU, CodeT5+\_LSTM, CodeT5+\_BiLSTM, CodeT5+\_BiLSTM-A, UniXcoder\_GRU, UniXcoder\_LSTM, UniXcoder\_BiLSTM, UniXcoder\_BiLSTM-A, RoBERTa\_GRU, RoBERTa\_LSTM, RoBERTa\_BiLSTM, RoBERTa\_BiLSTM-A, RoBERTa\_LR\_GRU, RoBERTa\_LR\_LSTM, RoBERTa\_LR\_BiLSTM, RoBERTa\_LR\_BiLSTM-A, PLBART\_GRU, PLBART\_LSTM, PLBART\_BiLSTM, PLBART\_BiLSTM-A, CoTexT\_GRU, CoTexT\_LSTM, CoTexT\_BiLSTM, CoTexT\_BiLSTM-A},
    xtick=data,
    x tick label style={rotate=25, anchor=east, inner sep=-2pt},
    nodes near coords,
    nodes near coords align={vertical},
    every node near coord/.append style={
        font=\scriptsize,
        anchor=west,
        rotate=80
    },
    legend style={
        font=\scriptsize,
        at={(0.5, -0.45)},
        anchor=north,
        legend columns=-1
    }
]
\addplot [color=blue, fill=blue!20, semithick, pattern=grid, pattern color = blue] coordinates {(CodeT5\_GRU,82.16) (CodeT5\_LSTM,81.96) (CodeT5\_BiLSTM,82.15) (CodeT5\_BiLSTM-A,79.99) (GraphCodeBERT\_GRU,81.98) (GraphCodeBERT\_LSTM,26.76) (GraphCodeBERT\_BiLSTM,81.54) (GraphCodeBERT\_BiLSTM-A,79.73) (CodeT5+\_GRU,82.43) (CodeT5+\_LSTM,42.94) (CodeT5+\_BiLSTM,81.98) (CodeT5+\_BiLSTM-A,79.19) (UniXcoder\_GRU,70.87) (UniXcoder\_LSTM,71.14) (UniXcoder\_BiLSTM,70.03) (UniXcoder\_BiLSTM-A,72.47) (RoBERTa\_GRU,80.18) (RoBERTa\_LSTM,15.03) (RoBERTa\_BiLSTM,26.47) (RoBERTa\_BiLSTM-A,15.03) (RoBERTa\_LR\_GRU,79.59) (RoBERTa\_LR\_LSTM,79.65) (RoBERTa\_LR\_BiLSTM,79.78) (RoBERTa\_LR\_BiLSTM-A,78.66) (PLBART\_GRU,78.25) (PLBART\_LSTM,78.87) (PLBART\_BiLSTM,78.35) (PLBART\_BiLSTM-A,78.29) (CoTexT\_GRU,79.70) (CoTexT\_LSTM,78.67) (CoTexT\_BiLSTM,79.31) (CoTexT\_BiLSTM-A,69.73)};  

\addplot [color=orange, fill=orange!20, semithick, pattern=crosshatch, pattern color = orange] coordinates {(CodeT5\_GRU,52.41) (CodeT5\_LSTM,52.23) (CodeT5\_BiLSTM,52.55) (CodeT5\_BiLSTM-A,49.64) (GraphCodeBERT\_GRU,53.19) (GraphCodeBERT\_LSTM,02.30) (GraphCodeBERT\_BiLSTM,52.85) (GraphCodeBERT\_BiLSTM-A,49.54) (CodeT5+\_GRU,53.78) (CodeT5+\_LSTM,11.05) (CodeT5+\_BiLSTM,52.79) (CodeT5+\_BiLSTM-A,49.07) (UniXcoder\_GRU,35.92) (UniXcoder\_LSTM,37.28) (UniXcoder\_BiLSTM,36.84) (UniXcoder\_BiLSTM-A,38.26) (RoBERTa\_GRU,50.29) (RoBERTa\_LSTM,07.80) (RoBERTa\_BiLSTM,09.98) (RoBERTa\_BiLSTM-A,07.80) (RoBERTa\_LR\_GRU,49.36) (RoBERTa\_LR\_LSTM,49.71) (RoBERTa\_LR\_BiLSTM,49.64) (RoBERTa\_LR\_BiLSTM-A,48.30) (PLBART\_GRU,45.80) (PLBART\_LSTM,46.61) (PLBART\_BiLSTM,45.84) (PLBART\_BiLSTM-A,45.16) (CoTexT\_GRU,46.98) (CoTexT\_LSTM,45.63) (CoTexT\_BiLSTM,46.80) (CoTexT\_BiLSTM-A,36.13)};  

\legend{$F1_{\psi}$, $EMAcc$}
\end{axis}

%% file: Graphs_Figures/ROC_OE.tex
\begin{axis}[
    name=roc_auc,
    width=16.1cm,
    height=3cm,
    ylabel={AUC$_\psi$},
    ylabel style={yshift=-0.09em, font=\scriptsize},
    ytick={0.45, 0.50, 0.55, 0.60, 0.65, 0.70, 0.75, 0.80, 0.85, 0.90, 0.94},
    yticklabel style={font=\scriptsize, /pgf/number format/fixed},
    xtick=data,
    xticklabels={CodeT5\_GRU, CodeT5\_LSTM, CodeT5\_BiLSTM, CodeT5\_BiLSTM-A, GraphCodeBERT\_GRU, GraphCodeBERT\_LSTM, GraphCodeBERT\_BiLSTM, GraphCodeBERT\_BiLSTM-A, CodeT5+\_GRU, CodeT5+\_LSTM, CodeT5+\_BiLSTM, CodeT5+\_BiLSTM-A, UniXcoder\_GRU, UniXcoder\_LSTM, UniXcoder\_BiLSTM, UniXcoder\_BiLSTM-A, RoBERTa\_GRU, RoBERTa\_LSTM, RoBERTa\_BiLSTM, RoBERTa\_BiLSTM-A, RoBERTa\_LR\_GRU, RoBERTa\_LR\_LSTM, RoBERTa\_LR\_BiLSTM, RoBERTa\_LR\_BiLSTM-A, PLBART\_GRU, PLBART\_LSTM, PLBART\_BiLSTM, PLBART\_BiLSTM-A, CoTexT\_GRU, CoTexT\_LSTM, CoTexT\_BiLSTM, CoTexT\_BiLSTM-A},
    x tick label style={rotate=30, anchor=east, font=\scriptsize},
    ymin=0.45, ymax=0.94, 
    xmin=-0.5, xmax=31.5, 
    ymajorgrids=true,
    axis y line*=left,
    axis x line*=bottom,
    scale only axis=true, 
    enlargelimits=false, 
    legend style={at={(0.5,-0.52)}, anchor=north, legend columns=-1, font=\scriptsize}
]
\addplot+[mark=*, color=blue, thick] coordinates {
    (0,0.9314) (1,0.9258) (2,0.9266) (3,0.9283) 
    (4,0.9237) (5,0.5000) (6,0.9219) (7,0.9248) 
    (8,0.9283) (9,0.6857) (10,0.9276) (11,0.9250) 
    (12,0.8851) (13,0.8901) (14,0.8862) (15,0.8905) 
    (16,0.9213) (17,0.5000) (18,0.4947) (19,0.5000) 
    (20,0.9174) (21,0.9178) (22,0.9186) (23,0.9157) 
    (24,0.9049) (25,0.9080) (26,0.9034) (27,0.9011) 
    (28,0.9115) (29,0.9173) (30,0.9143) (31,0.8820)
};
\addlegendentry{ROC-AUC$_\psi$}
\addplot[draw=none, mark=triangle*, color=brown] coordinates {(0,0)};
\addlegendentry{One-Error ($OE$)}
\end{axis}

\begin{axis}[
    at={(roc_auc.south west)}, 
    anchor=south west, 
    width=16.1cm,
    height=3cm,
    axis y line*=right,
    axis x line=none,
    xtick=data,
    xticklabels=\empty, 
    xlabel={}, 
    ylabel={$OE$},
    ylabel style={yshift=0.5em, font=\scriptsize},
    ytick={0.05, 0.10, 0.15, 0.20, 0.25, 0.30, 0.35, 0.40, 0.45},
    yticklabel style={font=\scriptsize, /pgf/number format/fixed},
    ymin=0.05, ymax=0.45,
    xmin=-0.5, xmax=31.5, 
    restrict x to domain=0:31, 
    ymajorgrids=true,
    scale only axis=true, 
    enlargelimits=false, 
    clip=false 
]
\addplot+[mark=triangle*, color=brown, thick] coordinates {
    (0,0.0748) (1,0.0742) (2,0.0710) (3,0.0788) 
    (4,0.0788) (5,0.4471) (6,0.0774) (7,0.0842) 
    (8,0.0708) (9,0.3860) (10,0.0722) (11,0.0893) 
    (12,0.1456) (13,0.1426) (14,0.1482) (15,0.1415) 
    (16,0.0926) (17,0.4471) (18,0.4471) (19,0.4471) 
    (20,0.0948) (21,0.0955) (22,0.0946) (23,0.0971) 
    (24,0.1131) (25,0.1009) (26,0.1099) (27,0.1124) 
    (28,0.0936) (29,0.0988) (30,0.0965) (31,0.1514)
};
\end{axis}

%% file: Graphs_Figures/AV_JS.tex
\begin{axis}[
    ybar,
    width=19.0cm,
    height=4.5cm,
    bar width=4pt,                                
    ymin=0,
    ymax=90,
    ylabel={Scores (\%)},
    enlarge x limits=0.04,                        
    enlarge y limits={0.1, upper},
    grid=both,
    ytick={0,10,...,90},
    tick label style={font=\scriptsize},
    ylabel near ticks,
    ylabel shift={-5pt},
    symbolic x coords={CodeT5\_GRU, CodeT5\_LSTM, CodeT5\_BiLSTM, CodeT5\_BiLSTM-A, GraphCodeBERT\_GRU, GraphCodeBERT\_LSTM, GraphCodeBERT\_BiLSTM, GraphCodeBERT\_BiLSTM-A, CodeT5+\_GRU, CodeT5+\_LSTM, CodeT5+\_BiLSTM, CodeT5+\_BiLSTM-A, UniXcoder\_GRU, UniXcoder\_LSTM, UniXcoder\_BiLSTM, UniXcoder\_BiLSTM-A, RoBERTa\_GRU, RoBERTa\_LSTM, RoBERTa\_BiLSTM, RoBERTa\_BiLSTM-A, RoBERTa\_LR\_GRU, RoBERTa\_LR\_LSTM, RoBERTa\_LR\_BiLSTM, RoBERTa\_LR\_BiLSTM-A, PLBART\_GRU, PLBART\_LSTM, PLBART\_BiLSTM, PLBART\_BiLSTM-A, CoTexT\_GRU, CoTexT\_LSTM, CoTexT\_BiLSTM, CoTexT\_BiLSTM-A},
    xtick=data,
    x tick label style={rotate=25, anchor=east, inner sep=-2pt},
    nodes near coords,
    nodes near coords align={vertical},
    every node near coord/.append style={
        font=\scriptsize,
        anchor=west,
        rotate=80
    },
    legend style={
        font=\scriptsize,
        at={(0.5, -0.43)},
        anchor=north,
        legend columns=-1
    }
]
\addplot [color=teal, fill=teal!20, semithick, pattern=crosshatch dots, pattern color = teal] coordinates {(CodeT5\_GRU,91.57) (CodeT5\_LSTM,91.55) (CodeT5\_BiLSTM,91.63) (CodeT5\_BiLSTM-A,91.16) (GraphCodeBERT\_GRU,91.58) (GraphCodeBERT\_LSTM,76.72) (GraphCodeBERT\_BiLSTM,91.61) (GraphCodeBERT\_BiLSTM-A,91.08) (CodeT5+\_GRU,91.84) (CodeT5+\_LSTM,79.55) (CodeT5+\_BiLSTM,91.67) (CodeT5+\_BiLSTM-A,90.88) (UniXcoder\_GRU,87.86) (UniXcoder\_LSTM,88.16) (UniXcoder\_BiLSTM,88.00) (UniXcoder\_BiLSTM-A,88.28) (RoBERTa\_GRU,90.97) (RoBERTa\_LSTM,77.16) (RoBERTa\_BiLSTM,76.59) (RoBERTa\_BiLSTM-A,77.16) (RoBERTa\_LR\_GRU,90.77) (RoBERTa\_LR\_LSTM,90.76) (RoBERTa\_LR\_BiLSTM,90.87) (RoBERTa\_LR\_BiLSTM-A,90.52) (PLBART\_GRU,90.01) (PLBART\_LSTM,90.23) (PLBART\_BiLSTM,90.09) (PLBART\_BiLSTM-A,89.88) (CoTexT\_GRU,90.58) (CoTexT\_LSTM,90.14) (CoTexT\_BiLSTM,90.34) (CoTexT\_BiLSTM-A,87.74)};  

\addplot [color=black, fill=black!20, semithick, pattern=horizontal lines, pattern color = black] coordinates {(CodeT5\_GRU,75.80) (CodeT5\_LSTM,75.83) (CodeT5\_BiLSTM,76.10) (CodeT5\_BiLSTM-A,73.28) (GraphCodeBERT\_GRU,75.88) (GraphCodeBERT\_LSTM,32.43) (GraphCodeBERT\_BiLSTM,75.56) (GraphCodeBERT\_BiLSTM-A,72.84) (CodeT5+\_GRU,76.34) (CodeT5+\_LSTM,38.02) (CodeT5+\_BiLSTM,75.91) (CodeT5+\_BiLSTM-A,72.31) (UniXcoder\_GRU,62.01) (UniXcoder\_LSTM,62.32) (UniXcoder\_BiLSTM,61.07) (UniXcoder\_BiLSTM-A,63.71) (RoBERTa\_GRU,73.54) (RoBERTa\_LSTM,24.89) (RoBERTa\_BiLSTM,33.67) (RoBERTa\_BiLSTM-A,24.89) (RoBERTa\_LR\_GRU,72.83) (RoBERTa\_LR\_LSTM,72.87) (RoBERTa\_LR\_BiLSTM,72.92) (RoBERTa\_LR\_BiLSTM-A,71.61) (PLBART\_GRU,70.63) (PLBART\_LSTM,71.39) (PLBART\_BiLSTM,70.90) (PLBART\_BiLSTM-A,70.77) (CoTexT\_GRU,72.17) (CoTexT\_LSTM,71.26) (CoTexT\_BiLSTM,71.98) (CoTexT\_BiLSTM-A,60.68)};  

\legend{$AvgAcc$, $J_s$}
\end{axis}

%% file: Graphs_Figures/PwHm.tex
\begin{axis}[
    width=18.5cm,
    height=4.5cm,
    ybar,
    bar width=3pt,
    ylabel={Score},
    ylabel style={font=\normalsize, yshift=0.01em},
    ymin=0, ymax=1.0,
    ytick={0,0.1,0.2,0.4,0.6,0.8,1.0},
    scaled y ticks=false,
     y tick label style={/pgf/number format/fixed, font=\scriptsize},
    enlarge x limits=0.02,
    axis lines=box,
    xtick=data,
    xticklabels={CodeT5\_GRU, CodeT5\_LSTM, CodeT5\_BiLSTM, CodeT5\_BiLSTM-A, GraphCodeBERT\_GRU, GraphCodeBERT\_LSTM, GraphCodeBERT\_BiLSTM, GraphCodeBERT\_BiLSTM-A, CodeT5+\_GRU, CodeT5+\_LSTM, CodeT5+\_BiLSTM, CodeT5+\_BiLSTM-A, UniXcoder\_GRU, UniXcoder\_LSTM, UniXcoder\_BiLSTM, UniXcoder\_BiLSTM-A, RoBERTa\_GRU, RoBERTa\_LSTM, RoBERTa\_BiLSTM, RoBERTa\_BiLSTM-A, RoBERTa\_LR\_GRU, RoBERTa\_LR\_LSTM, RoBERTa\_LR\_BiLSTM, RoBERTa\_LR\_BiLSTM-A, PLBART\_GRU, PLBART\_LSTM, PLBART\_BiLSTM, PLBART\_BiLSTM-A, CoTexT\_GRU, CoTexT\_LSTM, CoTexT\_BiLSTM, CoTexT\_BiLSTM-A},
    x tick label style={rotate=25, anchor=east, font=\scriptsize},
    legend style={at={(0.5,-0.48)}, anchor=north, legend columns=-1, font=\scriptsize},
    ymajorgrids=true
]
\addplot+[ybar, draw=green!50!black, fill=green!20, pattern=north east lines, pattern color=green!50!black,] coordinates {
    (0,0.8334) (1,0.8336) (2,0.8348) (3,0.8533)
    (4,0.8388) (5,0.2032) (6,0.8509) (7,0.8534)
    (8,0.8438) (9,0.5195) (10,0.8412) (11,0.8521)
    (12,0.7994) (13,0.8175) (14,0.8235) (15,0.8026)
    (16,0.8345) (17,0.1167) (18,0.2007) (19,0.1167)
    (20,0.8308) (21,0.8343) (22,0.8392) (23,0.8362)
    (24,0.8129) (25,0.8106) (26,0.8104) (27,0.7966)
    (28,0.8174) (29,0.8107) (30,0.8120) (31,0.8144)
};
\addlegendentry{Precision}

\addplot[
    thick,
    color=black,
    mark=diamond*,
    mark options={fill=red},
    sharp plot
] table [x expr=\coordindex, y index=1] {
0 0.0843
1 0.0845
2 0.0837
3 0.0884
4 0.0842
5 0.2328
6 0.0839
7 0.0892
8 0.0816
9 0.2045
10 0.0833
11 0.0912
12 0.1214
13 0.1184
14 0.1200
15 0.1172
16 0.0903
17 0.2284
18 0.2341
19 0.2284
20 0.0923
21 0.0924
22 0.0913
23 0.0948
24 0.0999
25 0.0977
26 0.0991
27 0.1012
28 0.0942
29 0.0986
30 0.0966
31 0.1226
};
\addlegendentry{Hamming Loss}
\end{axis}

%% file: Graphs_Figures/AmiRm.tex
\begin{axis}[
    width=7cm,
    height=4cm,
    ylabel={AUC${\mu}$ (\%)},
    ylabel style={font=\scriptsize, yshift=0.01em},
    ymin=0, ymax=100,
    ytick={0, 10, 20, 30, 40, 50, 60, 70, 80, 90, 100},
    xtick=data,
    xticklabel style={font=\scriptsize, rotate=25, anchor=east},
    symbolic x coords={CodeT5\_GRU, GraphCodeBERT\_BiLSTM-A, CodeT5+\_GRU, UniXcoder\_BiLSTM-A, RoBERTa\_GRU, RoBERTa\_LR\_BiLSTM, PLBART\_LSTM, CoTexT\_LSTM},
    yticklabel style={font=\scriptsize},
    axis y line*=left,
    axis x line*=bottom,
    legend style={at={(0.5,-0.48)}, anchor=north, legend columns=2, font=\scriptsize}
]
\addplot[
    color=violet,
    mark=star,
    thick,
    densely dashed
] coordinates {
    (CodeT5\_GRU, 95.58)
    (GraphCodeBERT\_BiLSTM-A, 95.32)
    (CodeT5+\_GRU, 95.46)
    (UniXcoder\_BiLSTM-A, 93.35)
    (RoBERTa\_GRU, 95.07)
    (RoBERTa\_LR\_BiLSTM, 94.90)
    (PLBART\_LSTM, 94.22)
    (CoTexT\_LSTM, 94.88)
}; 
\addlegendentry{ROC-AUC (${\mu}$)}
\end{axis}

\begin{axis}[
    width=7cm,
    height=4cm,
    ylabel={$R_{\bar{\mu}}$ (\%)},
    ylabel style={font=\scriptsize, yshift=0.01em},
    ymin=0, ymax=100,
    xtick=data,
    xticklabel style={font=\scriptsize, rotate=25, anchor=west},
    symbolic x coords={CodeT5\_LSTM, GraphCodeBERT\_GRU, CodeT5+\_GRU, UniXcoder\_BiLSTM-A, RoBERTa\_GRU, RoBERTa\_LR\_LSTM, PLBART\_LSTM, CoTexT\_GRU},
    yticklabels={,,}, 
    axis y line*=right,
    axis x line*=top,
    legend style={at={(0.5,1.7)}, anchor=north, legend columns=2, font=\scriptsize}
]
\addplot[
    color=olive,
    mark=+,
    thick,
    solid
] coordinates {
    (CodeT5\_LSTM, 75.09)
    (GraphCodeBERT\_GRU, 73.13)
    (CodeT5+\_GRU, 73.37)
    (UniXcoder\_BiLSTM-A, 53.43)
    (RoBERTa\_GRU, 68.59)
    (RoBERTa\_LR\_LSTM, 69.42)
    (PLBART\_LSTM, 69.88)
    (CoTexT\_GRU, 71.22)
}; 
\addlegendentry{Recall ($R_{\bar{\mu}}$)}
\end{axis}

%% file: Graphs_Figures/Time.tex
\pgfplotsset{compat=1.18}
\begin{axis}[
  width=18cm, height=4.5cm,
  ylabel={Time (minutes)},
  ymin=0, ymax=1500, ymajorgrids, grid=both,
  ytick={0,200,400,600,800,1000,1200,1400},
  xtick={1,...,32},
  xticklabel style={rotate=28, anchor=east, font=\scriptsize},
  tick label style={font=\small},
  enlargelimits=0.01,
  legend style={
    draw=none, at={(0.5,-0.46)}, anchor=north,
    legend columns=3, /tikz/every even column/.append style={column sep=0.8cm}
  },
  xticklabels={
    CodeT5\_GRU, CodeT5\_LSTM, CodeT5\_BiLSTM, CodeT5\_BiLSTM-A,
    GraphCodeBERT\_GRU, GraphCodeBERT\_LSTM, GraphCodeBERT\_BiLSTM, GraphCodeBERT\_BiLSTM-A,
    CodeT5+\_GRU, CodeT5+\_LSTM, CodeT5+\_BiLSTM, CodeT5+\_BiLSTM-A,
    UniXcoder\_GRU, UniXcoder\_LSTM, UniXcoder\_BiLSTM, UniXcoder\_BiLSTM-A,
    RoBERTa\_GRU, RoBERTa\_LSTM, RoBERTa\_BiLSTM, RoBERTa\_BiLSTM-A,
    RoBERTa\_LR\_GRU, RoBERTa\_LR\_LSTM, RoBERTa\_LR\_BiLSTM, RoBERTa\_LR\_BiLSTM-A,
    PLBART\_GRU, PLBART\_LSTM, PLBART\_BiLSTM, PLBART\_BiLSTM-A,
    CoTexT\_GRU, CoTexT\_LSTM, CoTexT\_BiLSTM, CoTexT\_BiLSTM-A
  }
]

\addplot+[very thick, color=lime!80!black, mark=x, mark size=2.5pt]
coordinates {
(1,224.04) (2,610.46) (3,240.93) (4,112.93)
(5,226.61) (6,231.91) (7,241.59) (8,125.79)
(9,225.92) (10,233.53) (11,231.87) (12,112.91)
(13,192.55) (14,93.71) (15,126.60) (16,100.59)
(17,245.21) (18,229.80) (19,225.14) (20,247.53)
(21,207.07) (22,214.87) (23,213.55) (24,223.43)
(25,180.70) (26,132.26) (27,170.56) (28,130.79)
(29,224.63) (30,220.05) (31,229.09) (32,100.22)
};
\addlegendentry{Optuna}

\addplot+[very thick, color=violet, mark=pentagon*, mark size=2.2pt]
coordinates {
(1,409.66) (2,695.32) (3,739.88) (4,249.60)
(5,394.09) (6,407.76) (7,491.65) (8,329.55)
(9,495.47) (10,528.48) (11,421.62) (12,803.46)
(13,232.42) (14,178.32) (15,186.95) (16,144.65)
(17,949.51) (18,348.85) (19,412.25) (20,507.11)
(21,474.14) (22,431.04) (23,360.93) (24,496.87)
(25,266.40) (26,523.48) (27,231.65) (28,234.65)
(29,406.90) (30,451.53) (31,420.39) (32,166.40)
};
\addlegendentry{Final Training}

\addplot+[very thick, color=teal, mark=+, mark size=2.5pt]
coordinates {
(1,633.70) (2,1305.78) (3,980.81) (4,362.53)
(5,620.69) (6,639.68) (7,733.24) (8,455.34)
(9,721.39) (10,762.01) (11,653.49) (12,916.37)
(13,424.97) (14,272.03) (15,313.55) (16,245.25)
(17,1194.72) (18,578.65) (19,637.39) (20,754.64)
(21,681.21) (22,645.91) (23,574.48) (24,720.30)
(25,447.10) (26,655.74) (27,402.21) (28,365.44)
(29,631.52) (30,671.58) (31,649.48) (32,266.62)
};
\addlegendentry{Total Time}

\end{axis}